\documentclass[11pt,a4paper]{article}
\pdfoutput=1
\usepackage{jcappub}
\usepackage{graphicx}
\usepackage{xcolor}
\usepackage{mathrsfs,mathtools}
\usepackage{physics,amssymb}
\usepackage{siunitx}
\usepackage{bm}
\usepackage[normalem]{ulem}
\usepackage{cancel}
\usepackage{url}
\usepackage{longtable}
\usepackage{xspace}
\usepackage{acronym}

\hypersetup{colorlinks=true
,urlcolor=DARKBLUE
,anchorcolor=DARKBLUE
,citecolor=DARKBLUE
,filecolor=DARKBLUE
,linkcolor=DARKBLUE
,menucolor=DARKBLUE
,linktocpage=true
,pdfproducer=medialab
,pdfa=true
}

\DeclareMathOperator{\sinc}{sinc}

\newcommand{\ee}{\mathrm{e}}
\newcommand{\kdx}{\mathbf{k}\cdot\mathbf{x}}
\newcommand{\Mpl}{M_\mathrm{Pl}}
\newcommand{\ns}{n_{\mathrm{s}}}

\newcommand{\IR}{\mathrm{IR}}
\newcommand{\UV}{\mathrm{UV}}
\newcommand{\cg}{\mathrm{cg}}
\newcommand{\cl}{\mathrm{cl}}
\newcommand{\umin}{\mathrm{min}}
\newcommand{\umax}{\mathrm{max}}
\newcommand{\fNL}{f_\mathrm{NL}}
\newcommand{\uth}{\mathrm{th}}
\newcommand{\PBH}{\mathrm{PBH}}
\newcommand{\DM}{\mathrm{DM}}
\newcommand{\peak}{\mathrm{peak}}
\newcommand{\MS}{\mathrm{MS}}

\newcommand{\calB}{\mathcal{B}}

\newcommand{\ub}{\mathrm{b}}
\newcommand{\calC}{\mathcal{C}}
\newcommand{\scrC}{\mathscr{C}}

\newcommand{\uD}{\mathrm{D}}

\newcommand{\uF}{\mathrm{F}}

\newcommand{\uG}{\mathrm{G}}

\newcommand{\sfH}{\mathsf{H}}
\newcommand{\uI}{\mathrm{I}}
\newcommand{\ui}{\mathrm{i}}

\newcommand{\bfk}{\mathbf{k}}
\newcommand{\calL}{\mathcal{L}}

\newcommand{\um}{\mathrm{m}}
\newcommand{\bfm}{\mathbf{m}}
\newcommand{\calN}{\mathcal{N}}
\newcommand{\scrN}{\mathscr{N}}
\newcommand{\frakN}{\mathfrak{N}}
\newcommand{\bfn}{\mathbf{n}}
\newcommand{\calO}{\mathcal{O}}

\newcommand{\uP}{\mathrm{P}}
\newcommand{\calP}{\mathcal{P}}

\newcommand{\scrR}{\mathscr{R}}

\newcommand{\us}{\mathrm{s}}

\newcommand{\ut}{\mathrm{t}}

\newcommand{\calW}{\mathcal{W}}
\newcommand{\bfx}{\mathbf{x}}
\newcommand{\bbZ}{\mathbb{Z}}

\newcommand{\epv}{\epsilon_{V}}
\newcommand{\hv}{\eta_{V}}
\newcommand{\bfzero}{\mathbf{0}}

\newcommand{\beae}[1]{\begin{equation}\begin{aligned} #1 \end{aligned}\end{equation}}

\newcommand{\bae}[1]{\begin{align} #1 \end{align}}
\newcommand{\bce}[1]{\begin{cases} #1 \end{cases}}

\newcommand{\bme}[1]{\begin{multline} #1 \end{multline}}
\newcommand{\bmte}[1]{\begin{multlined}[t] #1 \end{multlined}}

\newcommand{\relmiddle}[1]{\mathrel{}\middle#1\mathrel{}}

\definecolor{MONZA}{HTML}{CF000F}
\definecolor{DARKBLUE}{HTML}{00008b}
\definecolor{DARKMAGENTA}{HTML}{8b008b}

\acrodef{PBH}{primordial black hole}
\acrodef{BH}{black hole}
\acrodef{CMB}{cosmic microwave background}
\acrodef{FLRW}{Friedmann--Lema\^itre--Robertson--Walker}
\acrodef{STOLAS}{STOchastic LAttice Simulation}
\acrodef{ADM}{Arnowitt--Deser--Misner}
\acrodef{PDF}{probability density function}
\acrodef{USR}{ultra-slow-roll}
\acrodef{EoM}{equation of motion}
\newacroplural{EoM}{equations of motion}
\acrodef{DoF}{degree of freedom}
\newacroplural{DoF}{degrees of freedom}
\acrodef{RK4}{fourth-order Runge--Kutta}
\acrodef{PDF}{probability density function}
\acrodef{GW}{gravitational wave}
\acrodef{DM}{dark matter}
\acrodef{FFT}{Fast Fourier Transform}
\acrodef{DFT}{Discrete Fourier Transform}
\acrodef{NLO}{next-to-leading order}

\begin{document}
\title{STOLAS: STOchastic LAttice Simulation of cosmic inflation}
\date{\today}

\collaborationImg{\includegraphics[width=0.3\hsize]{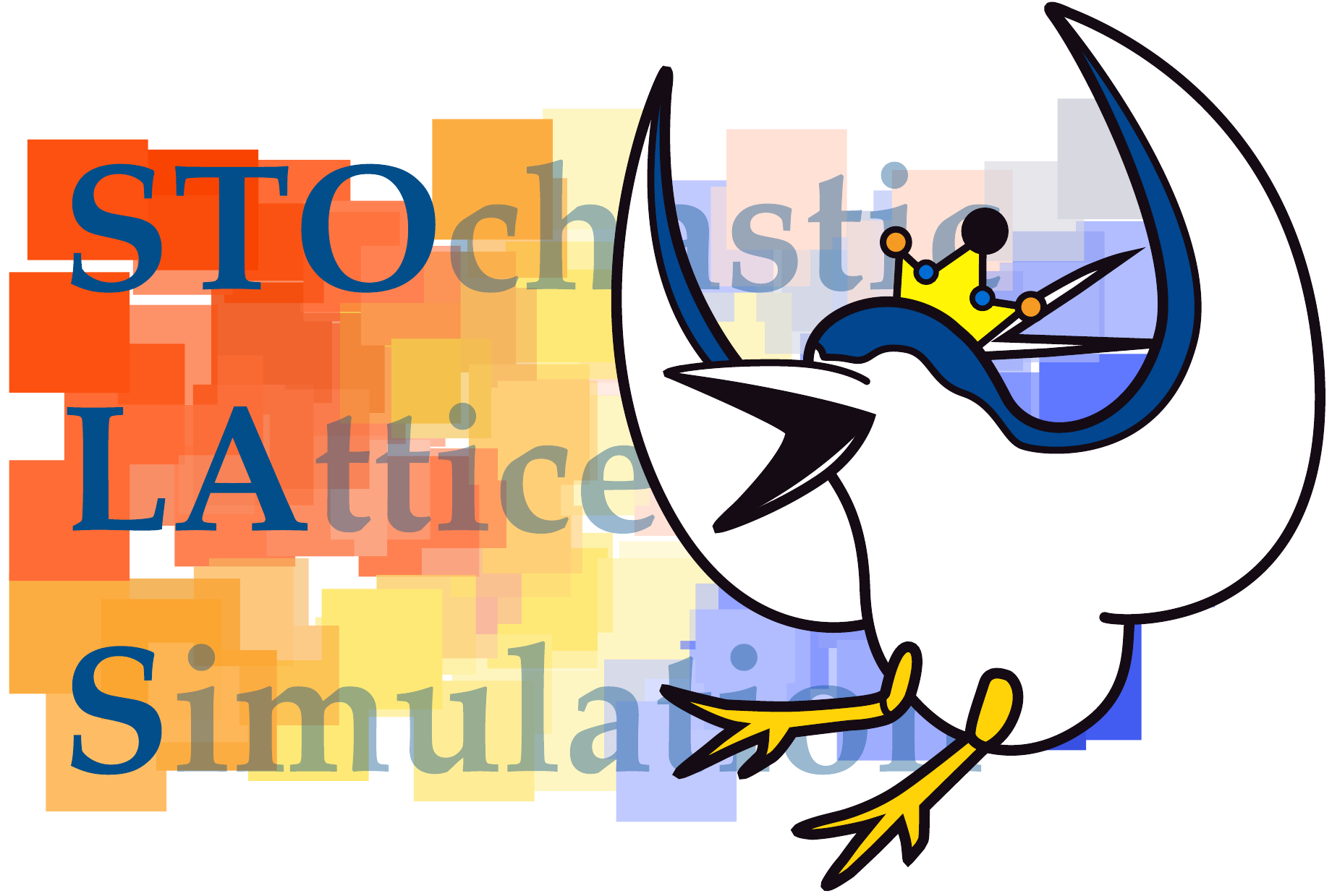}}
\author[a]{Yurino Mizuguchi,}
\author[b]{Tomoaki Murata,}
\author[c,a]{and Yuichiro Tada}

\affiliation[a]{Department of Physics, Nagoya University, \\
Furo-cho Chikusa-ku, Nagoya 464-8602, Japan}
\affiliation[b]{Department of Physics, Rikkyo University, \\ 
Toshima, Tokyo 171-8501, Japan}
\affiliation[c]{Institute for Advanced Research, Nagoya University, \\
Furo-cho Chikusa-ku, Nagoya 464-8601, Japan}

\emailAdd{mizuguchi.yurino.y0@s.mail.nagoya-u.ac.jp}
\emailAdd{tmurata@rikkyo.ac.jp}
\emailAdd{tada.yuichiro.y8@f.mail.nagoya-u.ac.jp}

\abstract{
We develop a C++ package of the \ac{STOLAS} of cosmic inflation.
It performs the numerical lattice simulation in the application of the stochastic-$\delta N$ formalism.
\ac{STOLAS} can directly compute the three-dimensional map of the observable curvature perturbation without estimating its statistical properties.
In its application to two toy models of inflation, chaotic inflation and Starobinsky's linear-potential inflation, we confirm that \ac{STOLAS} is well-consistent with the standard perturbation theory.
Furthermore, by introducing the importance sampling technique, we have success in numerically sampling the current abundance of \acp{PBH} in a non-perturbative way.
The package is available \href{https://github.com/STOchasticLAtticeSimulation/STOLAS_dist}{here}.
}
\arxivnumber{2405.10692}

\null\hfill\begin{tabular}[t]{l@{}}
  { RUP-24-10}
\end{tabular}

\maketitle

\acresetall
\section{Introduction}

Cosmic inflation, the accelerated expansion phase of the early universe, is a widely accepted scenario of the dawn of our universe.
Not only does it realise a globally flat and homogeneous universe, but inflation can also provide local fluctuations in the metric or the energy density by stretching the quantum vacuum fluctuation, those primordial perturbations growing into the current rich structures of the universe such as galaxies and clusters.
Since its advent~\cite{Starobinsky:1980te,Sato:1981qmu,Guth:1980zm,Linde:1981mu,Albrecht:1982wi,Linde:1983gd}, a tremendous number of inflation models have been proposed, but its concrete mechanism is yet to be clarified despite the significant improvement in its understanding through the cosmological observations represented by the Planck measurements~\cite{Planck:2018jri} of the \ac{CMB} anisotropies.
To figure out its whole aspect, further detailed studies on the properties of the generated primordial perturbation as well as the deep observations of cosmological structures as its descendants are necessary.

Speaking of the connection between the primordial perturbation and the late-time observables, the importance of numerical simulations such as $N$-body simulations for dark matter halos and hydrodynamic simulations for galaxy formations (see, e.g., Ref.~\cite{Vogelsberger:2019ynw} for a recent review on cosmological simulations) has been increasing more and more.
The lattice simulation of inflation~\cite{Caravano:2021pgc,Caravano:2021bfn,Caravano:2022epk,Caravano:2022yyv,Figueroa:2023oxc,Caravano:2024tlp} has started to attract attention but has been less developed compared to the simulations of the late universe.
One of the potential issues in the development of simulations of inflation is that quantum mechanics and the theory of gravity, the two principal elements of inflation, are unified only in a perturbative manner and no guidance principle for a non-perturbative simulation has been clarified yet.

The \emph{stochastic formalism} of inflation (see Refs.~\cite{Starobinsky:1982ee,Starobinsky:1986fx,Nambu:1987ef,Nambu:1988je,Kandrup:1988sc,Nakao:1988yi,Nambu:1989uf,Mollerach:1990zf,Linde:1993xx,Starobinsky:1994bd} for the first works and Ref.~\cite{Cruces:2022imf} for a recent review) can bring about a breakthrough in this situation.
It is an effective theory for superHubble coarse-grained fields (IR modes) which is obtained by integrating out the other UV modes perturbatively (and hence the quantum theory of gravity is well-defined at least at tree level).
The IR modes are well approximated by classical fields and the horizon exit of the UV modes are treated as classical stochastic noise onto the IR modes.
Furthermore, the gradient expansion (the long-wavelength approximation) justifies, at its lowest order, the assumption that the IR spacetime locally reads the flat \ac{FLRW} universe, which significantly simplifies the computation.
In this way, the stochastic formalism can provide a guidance principle for a numerical lattice simulation of the superHubble IR fields.
Coarse-graining in this formalism has a good compatibility with a lattice simulation as it anyway needs to ``discretise" the spacetime.

The lattice simulation of stochastic inflation has been done by Salopek and Bond~\cite{PhysRevD.43.1005} for the first time to simulate the inflaton field's fluctuation, but not for the observable curvature perturbation.
To calculate the curvature perturbation in the stochastic formalism, the \emph{stochastic-$\delta N$} technique has been intentionally developed in recent years (see Refs.~\cite{Fujita:2013cna,Fujita:2014tja,Vennin:2015hra} for the first papers and also Refs.~\cite{Figueroa:2020jkf,Figueroa:2021zah,Raatikainen:2023bzk} for applications to numerical simulations other than lattice).
In this work, we propose a C++ package, \textsc{\ac{STOLAS}}, for the curvature perturbation in combination with the stochastic-$\delta N$ formalism and the numerical lattice simulation.

The paper is organised as follows.
In Sec.~\ref{sec: Stochastic lattice simulation}, we review the stochastic formalism and provide the implementation of \ac{STOLAS}.
In Sec.~\ref{sec: Primary statistics}, we show that \ac{STOLAS} successfully reproduce the primary statistics such as the power spectrum and the non-linearity parameter of the curvature perturbation.
In Sec.~\ref{sec: Importance sampling}, we exemplify the calculation of the \ac{PBH} mass function as a powerful application of \ac{STOLAS}.
\acp{PBH} are the productions of $\calO(1)$ fluctuations which are good targets of the non-perturbative simulation.
Also, it has been recently claimed that the \ac{PBH} formation criterion is sensitive to the spatial profile of the large perturbation (see, e.g., Refs.~\cite{Atal:2019erb,Escriva:2019phb}), which can be simulated by the position-space lattice computation.
The \emph{importance sampling} technique (see Ref.~\cite{Jackson:2022unc} in the context of stochastic inflation) enables us to sample the \ac{PBH} formation, the statistically rare event.
Sec.~\ref{sec: Conclusions} is devoted to conclusions. We adopt the natural unit $c=\hbar=1$ throughout the paper.

\section{\acl{STOLAS}}\label{sec: Stochastic lattice simulation}

Our \ac{STOLAS} implements the stochastic formalism of inflation in the numerical lattice simulation.
Though the basic idea has already been proposed in 1991 by Salopek and Bond~\cite{PhysRevD.43.1005}, it has long been out of the spotlight.
The significant development in computational resources provides more opportunities for numerical simulations than in those days.
Furthermore, the recent stochastic-$\delta N$ technique~\cite{Fujita:2013cna,Fujita:2014tja,Vennin:2015hra} enables us to calculate not only the inflatons' fluctuations but also the observable curvature perturbations.
In this section, we first review the stochastic formalism of inflation and then describe its implementation in the discrete lattice simulation. 

\subsection{Stochastic formalism of inflation}

Let us review the stochastic formalism of inflation in this subsection.
First, we start from the action of general relativity and a canonical scalar field $\phi$:\footnote{See, e.g., Ref.~\cite{Pinol:2020cdp} for a generalisation to multiple scalar fields.} 
\begin{align}\label{action}
    S&=\int \dd[4]x\sqrt{-g}\qty[\frac{1}{2}\Mpl^2R-\frac{1}{2}g^{\mu\nu}\partial_\mu\phi\partial_\nu\phi-V(\phi)].
\end{align}
$R$ is the Ricci scalar associated with the spacetime metric $g_{\mu\nu}$, $V(\phi)$ is the scalar potential, and $\Mpl=1/\sqrt{8\pi G}$ is the reduced Planck mass.
We adopt the \ac{ADM} decomposition of the spacetime metric as
\bae{
    \dd{s^2}=-\scrN^2\dd{t^2}+\gamma_{ij}(\dd{x^i}+\beta^i\dd{t})(\dd{x^j}+\beta^j\dd{t}),
}
where $\scrN$ is the lapse function, $\beta^i$ is the shift vector, and $\gamma_{ij}$ is the spatial metric.
The corresponding Lagrangian density $S=\int\dd{t}\dd[3]{x}\calL$ reads
\bae{
    \calL=\scrN\sqrt{\gamma}\bqty{\frac{\Mpl^2}{2}\pqty{R^{(3)}+K_{ij}K^{ij}-K^2}+\frac{1}{2\scrN^2}v^2-\frac{1}{2}\gamma^{ij}\partial_i\phi\partial_j\phi-V(\phi)}.
}
Here, $\gamma=\det\gamma_{ij}$, $R^{(3)}$ is the corresponding spatial Ricci curvature,
\bae{
    K_{ij}=\frac{1}{2\scrN}\pqty{2\beta_{(i|j)}-\dot{\gamma}_{ij}},
}
is the extrinsic curvature (where dots denote time derivatives, the symbol $|$ indicates the covariant derivative associated with $\gamma_{ij}$, and the parentheses signal symmetrisation), $K=K^i_i$ is its trace where the spacial indices are raised and lowered by $\gamma_{ij}$ and its inverse $\gamma^{ij}$, and
\bae{
    v=\dot{\phi}-\beta^i\partial_i\phi.
}
This Lagrangian is followed by the constraint equations
\bae{\label{eq: constraint}
    C=C_i=0,
}
with
\beae{
    &C=\frac{2}{\gamma\Mpl^2}\bqty{\pi_{ij}\pi^{ij}-\frac{1}{2}\pqty{\pi^i_i}^2}-\frac{\Mpl^2}{2}R^{(3)}+\frac{1}{2\gamma}\pi_\uI^2+\frac{\gamma^{ij}}{2}\partial_i\phi\partial_j\phi+V, \\
    &C_i=-2\pqty{\frac{\pi^j_i}{\sqrt{\gamma}}}_{|j}+\frac{1}{\sqrt{\gamma}}\pi_\uI\partial_i\phi=\frac{1}{\sqrt{\gamma}}\pqty{-2\partial_k(\gamma_{ij}\pi^{jk})+\pi^{jk}\partial_i\gamma_{jk}+\pi_\uI\partial_i\phi},
}
and \acp{EoM} (in the scalar sector)
\beae{\label{eq: EoM}
    &\dot{\phi}=\frac{\scrN}{\sqrt{\gamma}}\pi_\uI+\beta^i\partial_i\phi, \\
    &\dot{\pi}_\uI=-\sqrt{\gamma}\scrN V'+\partial_i\pqty{\sqrt{\gamma}\scrN\gamma^{ij}\partial_j\phi}+\partial_i(\beta^i\pi_\uI),
}
with the conjugate momenta (the subscript ``$\uI$" stands for ``inflaton")
\bae{
    \pi_\uI=\pdv{\calL}{\dot{\phi}}=\frac{\sqrt{\gamma}}{\scrN}v \qc
    \pi^{ij}=\pdv{\calL}{\dot{\gamma}_{ij}}=\frac{\Mpl^2}{2}\sqrt{\gamma}(K\gamma^{ij}-K^{ij}).
}

For convenience (and conceptual relevance; see, e.g., Ref.~\cite{Pattison:2019hef}), we employ the e-folding number $N$ as the time variable and take the spatially flat gauge, neglecting the vector and tensor perturbations:
\bae{
    \gamma_{ij}(N,\bfx)=a^2(N)\delta_{ij} \qc
    a(N)\propto\ee^N,
}
where the scale factor $a(N)$ is spatially homogeneous. Relevant equations are simplified in this gauge as $R^{(3)}$ vanishes and $\sqrt{\gamma}=a^3$.
Hereafter, we also rescale the inflaton momentum as $\pi_\uI\to a^3\pi_\uI$.

In stochastic inflation, the physical variables (represented by the symbol $X$) are decomposed into the superHubble ``IR" part and the other ``UV" part as
\beae{\label{eq: XIR}
    &X_\IR(N,\bfx)=\int\frac{\dd[3]{k}}{(2\pi)^3}\ee^{i\kdx}X_\bfk(N)\Theta(\sigma a(N)\sfH-k), \\
    &X_\UV(N,\bfx)=X(N,\bfx)-X_\IR(N,\bfx), \\
}
where $\Theta(z)$ is the Heaviside step function,\footnote{The consistent definition of $\Theta(z)$ at $z=0$ requires a detailed discussion of the discretisation of the path integral. See Refs.~\cite{Tokuda:2017fdh,Tokuda:2018eqs}.}
\bae{
    \Theta(z)=\bce{
        0 & \text{for $z<0$,} \\
        1 & \text{for $z>0$}.
    }
}
The dimensionless and dimensionful parameters $\sigma$ and $\sfH$ are chosen so that $k_\sigma\coloneqq\sigma a(N)\sfH\ll a(N)H(N,\bfx)$ for relevant spacetime points and realisations, where the local Hubble parameter $H(N,\bfx)$ is defined below.\footnote{The literature often chooses $\sigma a(N)H(N,\bfx)$ with a positive small parameter $\sigma\ll1$ as the IR cutoff scale. Though the Hubble parameter is practically almost constant, this cutoff is spatial-dependent and circularly defined as the local Hubble parameter itself is defined by the IR quantities, strictly speaking. Hence, the consistent definition of the stochastic formalism in this way is tough and non-trivial. Our choice of the cutoff seems less physical but can avoid all these problems. From the viewpoint of the lattice simulation, $\sfH$ can be understood as a parameter of the grid size and causes no problem as long as $\sigma a(N)\sfH\ll a(N)H(N,\bfx)$.}
Practically, we will fix $\sfH$ by the global Hubble parameter at the simulation starting time and $\sigma$ by a positive and small number, $\sigma\ll1$.

We characterise the decomposed variables in our setup as
\beae{
    &\phi=\varphi+Q,\, &&\pi_\uI=\varpi+P, \\
    &\scrN=\scrN_\IR+\alpha,\, &&\beta^i=a^{-2}\delta^{ij}\partial_j\psi.
}
$\varphi$, $\varpi$, and $\scrN_\IR$ are IR while $Q$, $P$, $\alpha$, and $\psi$ are UV. We fixed $\beta_\IR^i=0$ as it is a pure gauge choice in the long-wavelength limit.\footnote{The long-wavelength shift may be non-local and affect the formulation of the stochastic equation for the non-attractor models. See, e.g., Ref.~\cite{Cruces:2021iwq} for the details.}
Substituting them into the constraints~\eqref{eq: constraint} and \acp{EoM}~\eqref{eq: EoM}, keeping terms at all orders in IR but up to linear order in UV, and dropping the spatial derivatives of IR quantities, one finds
\beae{\label{eq: IR EoM}
    &\dv{\varphi(N,\bfx)}{N}=\frac{\varpi(N,\bfx)}{H(N,\bfx)}+\xi_\phi(N,\bfx), \\
    &\dv{\varpi(N,\bfx)}{N}=-3\varpi(N,\bfx)-\frac{V'\pqty{\varphi(N,\bfx)}}{H(N,\bfx)}+\xi_\pi(N,\bfx),
}
for the IR part and the linearised equations
\beae{
    &\dv{Q_\bfk}{N}=\frac{P_\bfk}{H}+\frac{\varpi^2}{2\Mpl^2H^2}Q_\bfk, \\
    &\dv{P_\bfk}{N}=-3P_\bfk-\frac{k^2}{a^2H}Q_\bfk-\frac{1}{H}\pqty{V''+\frac{V'\varpi}{\Mpl^2H}+\frac{3\varpi^2}{2\Mpl^2}}Q_\bfk-\frac{\varpi^2}{2\Mpl^2H^2}P_\bfk,
}
for the UV part, which can be summarised into the standard Mukhanov--Sasaki equation~\eqref{eq: MSeq} (see, e.g., Ref.~\cite{Pinol:2020cdp}). The local Hubble parameter is defined by $H=1/\scrN_\IR$ and calculated via the local Friedmann equation
\bae{
    \frac{3\Mpl^2}{\scrN_\IR^2(N,\bfx)}=3\Mpl^2H^2(N,\bfx)=\frac{\varpi^2(N,\bfx)}{2}+V\pqty{\varphi(N,\bfx)},
}
derived from the constraint equation~\eqref{eq: constraint}.

$\xi$'s in the IR \ac{EoM}~\eqref{eq: IR EoM} are defined by
\beae{
    &\xi_\phi(N,\bfx)=\int\frac{\dd[3]{k}}{(2\pi)^3}\ee^{i\kdx}Q_\bfk(N)\partial_N\Theta(\sigma a(N)\sfH-k), \\
    &\xi_\pi(N,\bfx)=\int\frac{\dd[3]{k}}{(2\pi)^3}\ee^{i\kdx}P_\bfk(N)\partial_N\Theta(\sigma a(N)\sfH-k),
}
as a consequence of the time-dependent cutoff $\sigma a(N)\sfH$. Physically, they represent the horizon exit of subHubble modes. While the UV mode is understood as a quantum vacuum fluctuation, the superHubble IR mode is well approximated as a classical field. The $\xi$ terms hence have an intermediate feature: classical but random variables originating from the quantum vacuum fluctuation. Their statistical nature should be determined by the quantum theoretical expectation as (see, e.g., Ref.~\cite{Starobinsky:1994bd} for the detailed derivation)\footnote{$\xi_\phi$ and $\xi_\pi$ should be real because $\varphi$ and $\varpi$ are. However, the cross power spectrum $\calP_{\phi\pi}$ is not necessarily real in the definition~\eqref{eq: calP}. It exhibits the limitation of the ``heuristic" derivation of the stochastic formalism here. In a more sophisticated path-integral approach, one finds $\expval{\xi_\phi\xi_\pi}=\expval{\xi_\pi\xi_\phi}\propto\Re\calP_{\phi\pi}$; see Ref.~\cite{Pinol:2020cdp}. In this paper, we merely neglect $\xi_\pi$ hereafter and hence this subtlety is not problematic.}
\bae{\label{eq: xi correlation}
    \expval{\xi_X}=0 \qc
    \expval{\xi_X(N,\bfx)\xi_Y(N',\bfx')}=\calP_{XY}\pqty{N,k_\sigma(N)}\frac{\sin k_\sigma r}{k_\sigma r}\delta(N-N'), 
}
where $X$ and $Y$ represent $\phi$ or $\pi$, $r=\abs{\bfx-\bfx'}$ is the distance between $\bfx$ and $\bfx'$, and the dimensionless power spectrum $\calP_{XY}(N,k)$ is defined by
\bae{\label{eq: calP}
    \expval{Q_{X,\bfk}(N)Q_{Y,\bfk'}(N)}=(2\pi)^3\delta^{(3)}(\bfk+\bfk')\frac{2\pi^2}{k^3}\calP_{XY}(N,k),
    }
in the notation $Q_{\phi,\bfk}(N)=Q_\bfk(N)$ and $Q_{\pi,\bfk}(N)=P_\bfk(N)$. 
The braket in Eq.~\eqref{eq: calP} denote the quantum average, while other brakets in Eq.~\eqref{eq: xi correlation} and hereafter stand for the stochastic average.
Practically, the momentum noise is suppressed as $\calP_{\phi\pi}\sim\calO(\sigma^2)$ and $\calP_{\pi\pi}\sim\calO(\sigma^4)$ in a single-field slow-roll model.\footnote{Beyond this assumption such as the \ac{USR} models, the momentum noise can be non-negligible (see, e.g., Refs.~\cite{Pi:2022ysn,Kawaguchi:2023mgk,Jackson:2023obv}), though its implementation in \ac{STOLAS} is straightforward. We have confirmed that the power spectra in our two examples in Secs.~\ref{sec: Primary statistics} and \ref{sec: Importance sampling} are well consistent with the prediction of the Mukhanov--Sasaki equation even without the momentum noise, particularly around the peak of the power spectrum in Starobinsky's linear-potential model. The momentum noise could be relevant around the dip of the power spectrum in the \ac{USR} models~\cite{Dip}.}
We hence neglect $\xi_\pi$ and adopt the approximated \ac{EoM},
\beae{\label{eq: single noise EoM}
    &\dv{\varphi(N,\bfx)}{N}\simeq\frac{\varpi(N,\bfx)}{H(N,\bfx)}+\calP_\phi^{1/2}\pqty{N,k_\sigma(N)}\xi(N,\bfx), \\
    &\dv{\varpi(N,\bfx)}{N}\simeq-3\varpi(N,\bfx)-\frac{V'\pqty{\varphi(N,\bfx)}}{H(N,\bfx)},
}
with the renormalised noise,
\bae{
    \expval{\xi}=0 \qc 
    \expval{\xi(N,\bfx)\xi(N',\bfx')}=\frac{\sin k_\sigma r}{k_\sigma r}\delta(N-N').
}

\subsection{Discretisation}\label{sec: Discretisation}

In the numerical lattice simulation, we solve the dynamics of the stochastic inflation in the discretised spacetime with the time step $\Delta N$ and the grid size $\Delta x$ within the comoving box $L$ on a side. $N_L=\frac{L}{\Delta x}+1$ is the number of grids per side. We will take $N_L=64$ and our formulation will be based on the assumption that $N_L$ is an even number (a power of two, more specifically, to utilise the fast Fourier transformation algorithm).

In the simplest Euler--Maruyama method (but the application to a higher-order method is straightforward; see, e.g., Ref.~\cite{kloeden2011numerical}), the \ac{EoM}~\eqref{eq: single noise EoM} is discretised as\footnote{Note that the stochastic differential equation should be discretised in It\^o's way in the stochastic inflation (see Refs.~\cite{Tokuda:2017fdh,Tokuda:2018eqs}). The noise coefficient $\calP_\phi$ is hence evaluated at the time $N$ for the increment at $N$.
It\^o's discretisation generally breaks the field-space covariance~\cite{Pinol:2018euk}. Hence, in the curved-field-space multi-field models, the derivative should be replaced by the It\^o covariant derivative to keep the field-space covariance. See Ref.~\cite{Pinol:2020cdp} for the details.}
\beae{\label{eq: discrete EoM}
    &\varphi(N+\Delta N,\bfx)-\varphi(N,\bfx)=\Delta\varphi_\uD(N,\bfx)+\calP_\phi^{1/2}\pqty{N,k_\sigma(N)}\Delta W(N,\bfx), \\
    &\varpi(N+\Delta N,\bfx)-\varpi(N,\bfx)=\Delta\varpi_\uD(N,\bfx)
}
with the differences in the deterministic part
\bae{
    \Delta\varphi_\uD(N,\bfx)=\frac{\varpi(N,\bfx)}{H(N,\bfx)}\Delta N \qc
    \Delta\varpi_\uD(N,\bfx)=\pqty{-3\varpi(N,\bfx)-\frac{V'\pqty{\varphi(N,\bfx)}}{H(N,\bfx)}}\Delta N,
}
and the Gaussian random variable $\Delta W$ satisfying
\bae{\label{eq: dW correlation}
    \expval{\Delta W}=0 \qc 
    \expval{\Delta W(N,\bfx)\Delta W(N',\bfx')}=\frac{\sin k_\sigma r}{k_\sigma r}\delta_{NN'}\Delta N.
}
For stability, we adopt the \ac{RK4} method to evaluate the deterministic part $\Delta\varphi_\uD$ and $\Delta\varpi_\uD$.

The spatial dependence~\eqref{eq: dW correlation} can be realised with the use of the discrete Fourier space. We define the discrete Fourier and inverse Fourier transformation by
\bae{\label{eq: DFT of dW}
    \Delta W_\bfn(N)=\sum_\bfx\Delta W(N,\bfx)\ee^{-i\frac{2\pi}{L}\bfn\cdot\bfx} \qc
    \Delta W(N,\bfx)=\frac{1}{N_L^3}\sum_\bfn\Delta W_\bfn(N)\ee^{i\frac{2\pi}{L}\bfn\cdot\bfx},
}
where $\bfn=(n_x,n_y,n_z),\, (n_i\in\Bqty{0,1,2,\cdots,N_L-1})$ is the wave vector.
Note that the discrete Fourier mode exhibits the following periodicity:
\bae{\label{eq: periodicity}
    \Delta W_{(n_x,n_y,n_z)}=\Delta W_{(n_x+L,n_y,n_z)}=\Delta W_{(n_x,n_y+L,n_z)}=\Delta W_{(n_x,n_y,n_z+L)}.
}
It enables us to shift the wave vector domain to $\tilde{\bfn}=(\tilde{n}_x,\tilde{n}_y,\tilde{n}_z),\, \qty(\tilde{n}_i\in\Bqty{-\frac{N_L}{2}+1,-\frac{N_L}{2}+2,\cdots,\frac{N_L}{2}})$ with the definition
\bae{
    \tilde{n}_i=\bce{
        n_i & \text{for $n_i\leq N_L/2$}, \\
        n_i-N_L & \text{otherwise},
    }
    \quad\Leftrightarrow\quad
    n_i=\bce{
        \tilde{n}_i & \text{for $\tilde{n}_i\geq0$}, \\
        \tilde{n}_i+N_L & \text{for otherwise}.
    }
}
The original indexation is conventional for the fast Fourier transformation, while the reality condition in Fourier space, 
\bae{\label{eq: reality}
	\Delta W_{\tilde{\bfn}}=\Delta W_{-\tilde{\bfn}}^*
} 
is easy to handle in the shifted one.

Due to the reality condition, only the half of $\Delta W_{\bfn}$ is independent.
We specify the set $\scrC$ of the independent points by
\bae{
    \scrC=\bigcup_{i=1}^5\scrC_i,
}
where
\beae{
    &\scrC_1=\left\{\bfn\relmiddle{|}1\leq \tilde{n}_x\leq\frac{N_L}{2}-1,\, \tilde{n}_y\neq\frac{N_L}{2},\, \tilde{n}_z\neq \frac{N_L}{2}\right\}, \\
    &\scrC_2=\left\{\bfn\relmiddle{|}\tilde{n}_x=\frac{N_L}{2},\, \tilde{n}_y\neq \frac{N_L}{2},\, 1\leq \tilde{n}_z\leq\frac{N_L}{2}-1\right\}
    \cup \text{(2 even perms.)}, \\
    &\scrC_3=\left\{\bfn\relmiddle{|}\tilde{n}_x=0,\,\tilde{n}_y\neq\frac{N_L}{2},\,1\leq\tilde{n}_z\leq\frac{N_L}{2}-1\right\}, \\
    &\scrC_4=\left\{\bfn\relmiddle{|}\tilde{n}_x=\tilde{n}_y=\frac{N_L}{2},\,1\leq\tilde{n}_z\leq\frac{N_L}{2}-1\right\}\cup \text{(2 perms.)}, \\
    &\scrC_5=\bmte{\left\{\bfn\relmiddle{|}\tilde{n}_x=\tilde{n}_z=0,\,1\leq\tilde{n}_y\leq\frac{N_L}{2}-1\right\} \\
    \cup\bqty{\left\{\bfn\relmiddle{|}\tilde{n}_x=\frac{N_L}{2},\,1\leq\tilde{n}_y\leq\frac{N_L}{2}-1,\,\tilde{n}_z=0\right\}\cup \text{(2 even perms.)}}.}
}
Each set $\scrC_i$ includes the points of $(N_L/2-1)(N_L-1)^2$, $3(N_L/2-1)(N_L-1)$, $(N_L/2-1)(N_L-1)$, $3(N_L/2-1)$, and $4(N_L/2-1)$, reaching $(N_L^3-8)/2$ points in total. As they are complex, they correspond to $(N_L^3-8)$ real \acp{DoF}.
The remaining eight \acp{DoF} correspond to the following eight points:
\bae{
	\scrR=\left\{\bfn\relmiddle{|}\tilde{n}_i=0 \text{ or } \frac{N_L}{2} \quad (i=x,y,z)\right\}.
}
Their reflecting points are congruent to the original points modulo $N_L$:
\bae{
	\tilde{\bfn}-(-\tilde{\bfn})=2\tilde{\bfn}=(nN_L,mN_L,kN_L) \qc n,m,k\in\bbZ \qc \text{for $\tilde{\bfn}\in\scrR$}.
}
It hence follows from the periodicity~\eqref{eq: periodicity} and the reality condition~\eqref{eq: reality} that $\Delta W_\bfn$ for $\bfn\in\scrR$ is real.
The $N_L^3$ real \acp{DoF} expected in the position space are exhausted in total.

Finally, the spatial correlation~\eqref{eq: dW correlation} can be realised if $\Delta W_\bfn$ is non-zero only for $2\pi\tilde{n}/L=k_\sigma$.
Practically, we allow a small error for this condition, taking account of the discreteness of $\tilde{n}$ itself, as
\bae{
    \varsigma(N)=\left\{\bfn\relmiddle{|}\abs{\tilde{n}-n_\sigma(N)}\leq\frac{\Delta n_\sigma}{2}\right\} \qc
    n_\sigma(N)=\frac{k_\sigma(N)L}{2\pi},
}
where we choose $\Delta n_\sigma=1$.
The noise distribution is then determined at each time step by 
\bae{\label{eq: random_point}
    \bce{
        \Re[\Delta W_\bfn(N)]\sim\Im[\Delta W_\bfn(N)]\sim \frakN\pqty{0,\frac{N_L^6}{2\abs{\varsigma(N)}}\Delta N} & \text{for $\bfn\in\scrC\cap\varsigma(N)$}, \\
        \Delta W_\bfn(N)\sim \frakN\pqty{0,\frac{N_L^6}{\abs{\varsigma(N)}}\Delta N} & \text{for $\bfn\in\scrR\cap\varsigma(N)$}, \\
        \Delta W_\bfn(N)=\Delta W_{\bar{\bfn}}^* & \text{for $\bfn\in\overline{(\scrC\cup\scrR)}\cap\varsigma(N)$}, \\
        \Delta W_\bfn(N)=0 & \text{otherwise},
    }
}
where $\frakN(\mu,\sigma^2)$ is the normal distribution with the mean $\mu$ and the variance $\sigma^2$. 
Here, the symbol ``$\sim$" means that the random variables on both sides follow the same probability distribution, while the equal sign ``$=$" is used only if they take the same values.
$\abs{\varsigma(N)}$ is the number of elements of $\varsigma(N)$, and $\bar{\bfn}$ is the grid point congruent to $-\bfn$ modulo $N_L$:
\bae{
    \bar{n}_i=\bce{
        0 & \text{for $n_i=0$}, \\
        N_L-n_i & \text{otherwise}.
    }
}
With this noise distribution, the spatial correlation is reproduced as 
\bae{
    \expval{\Delta W(N,\bfx)\Delta W(N',\bfx')}&=\frac{\delta_{NN'}}{N_L^6}\sum_{\bfn,\bfm}\expval{\Delta W_\bfn(N)\Delta W_\bfm(N)}\ee^{i\frac{2\pi}{L}(\bfn\cdot\bfx+\bfm\cdot\bfx')} \nonumber \\
    &=\frac{\delta_{NN'}\Delta N}{\abs{\varsigma(N)}}\bqty{\sum_{\smqty{\bfn\in\scrC\cap\varsigma(N), \\
    \bfn\in\overline{(\scrC\cup\scrR)}\cap\varsigma(N)}}\ee^{i\frac{2\pi}{L}(\bfn\cdot\bfx+\bar{\bfn}\cdot\bfx')}+\sum_{\bfn\in\scrR\cap\varsigma(N)}\ee^{i\frac{2\pi}{L}\bfn\cdot(\bfx+\bfx')}} \nonumber \\
    &=\frac{\delta_{NN'}\Delta N}{\abs{\varsigma(N)}}\sum_{\bfn\in\varsigma(N)}\ee^{i\frac{2\pi}{L}\bfn\cdot(\bfx-\bfx')} \nonumber \\
    &\to\frac{\delta_{NN'}\Delta N}{4\pi}\int\dd{\Omega}\ee^{i\bfk_\sigma(N)\cdot(\bfx-\bfx')}=\frac{\sin k_\sigma(N)r}{k_\sigma(N)r}\delta_{NN'}\Delta N.
}
Note that $\exp[i\frac{2\pi}{L}\bar{\bfn}\cdot\bfx']=\exp[-i\frac{2\pi}{L}\bfn\cdot\bfx']$ for $\bfn\in\scrC$ and $\exp[i\frac{2\pi}{L}\bfn\cdot\bfx']=\exp[-i\frac{2\pi}{L}\bfn\cdot\bfx']$ for $\bfn\in\scrR$.
We also used the continuous limit in the last line.

\begin{figure}
	\centering
	\begin{tabular}{ccc}
        \begin{minipage}{0.3\hsize}
			\centering
			\includegraphics[width=0.9\hsize]{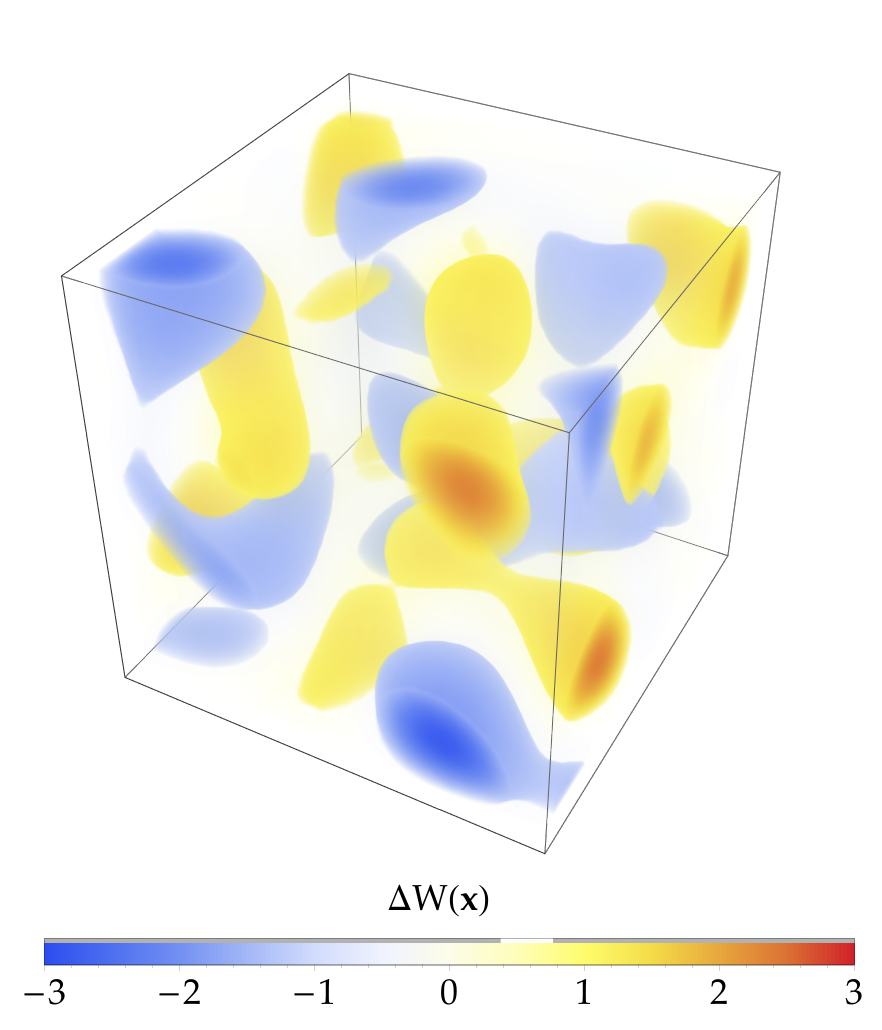}
		\end{minipage}&
		\begin{minipage}{0.3\hsize}
			\centering
			\includegraphics[width=0.9\hsize]{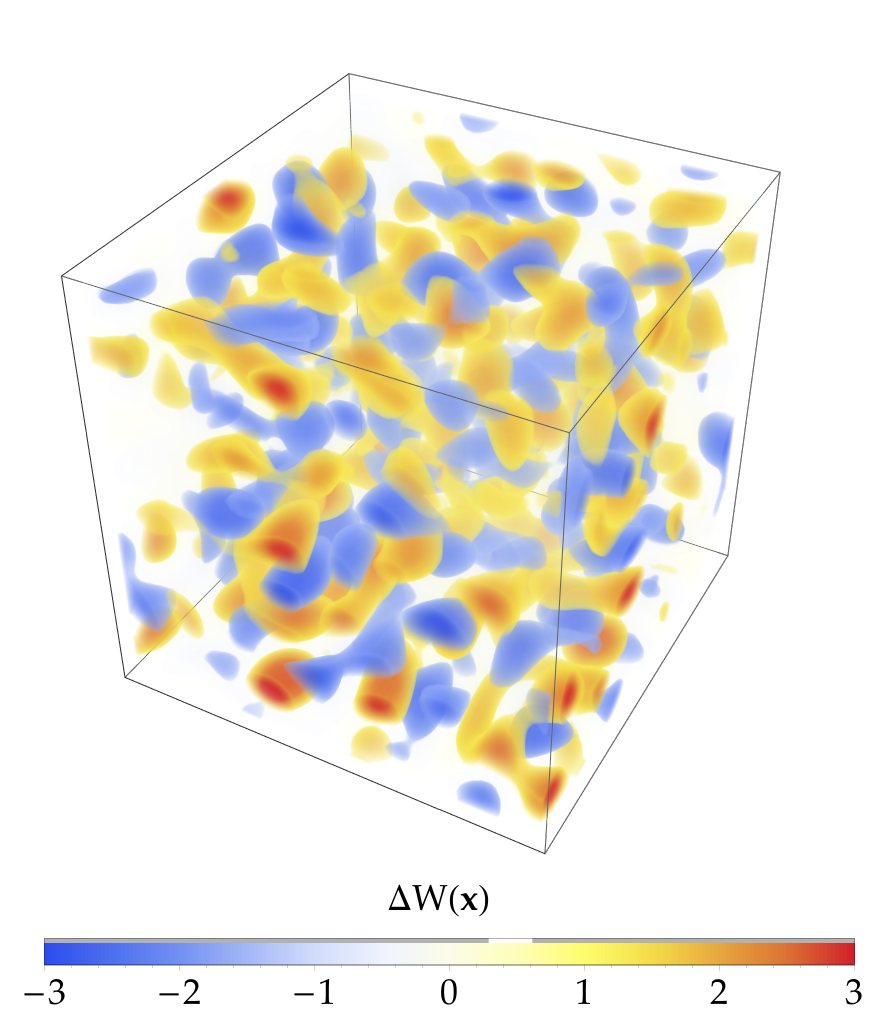}
		\end{minipage}&
    	\begin{minipage}{0.3\hsize}
			\centering
			\includegraphics[width=0.9\hsize]{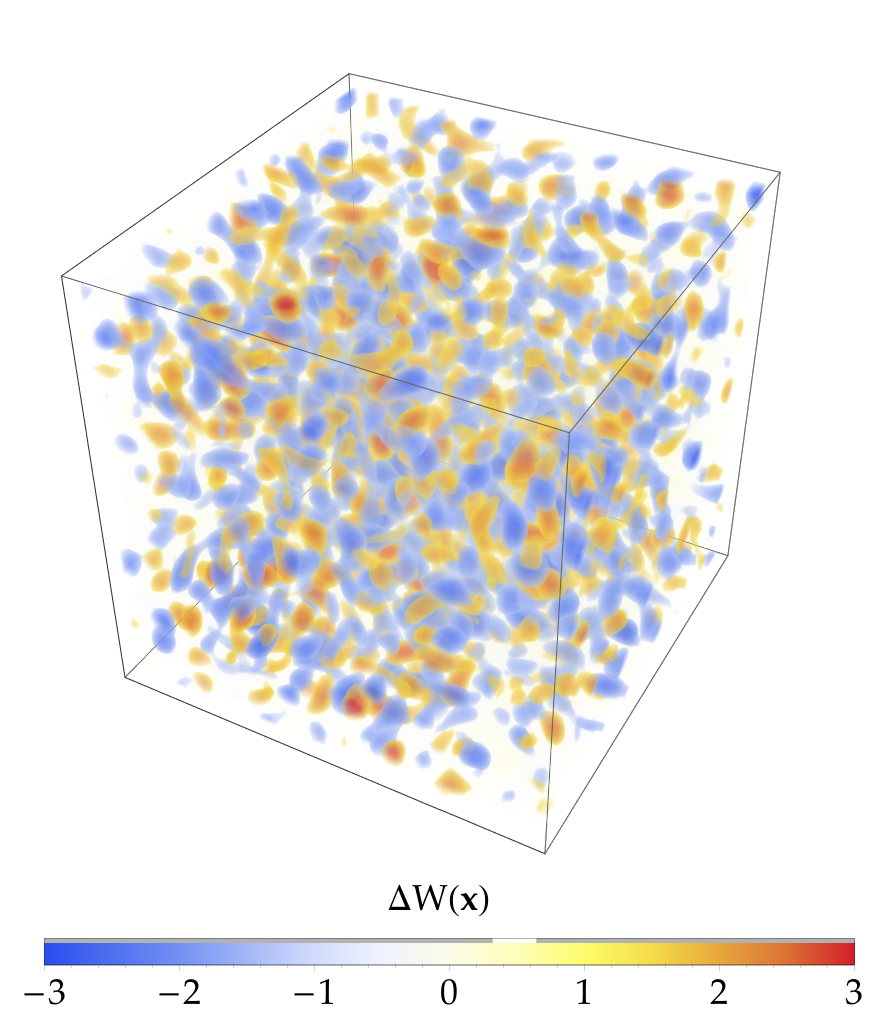}
		\end{minipage}\\
        \\
    	\begin{minipage}{0.3\hsize}
			\centering
			\includegraphics[width=\hsize]{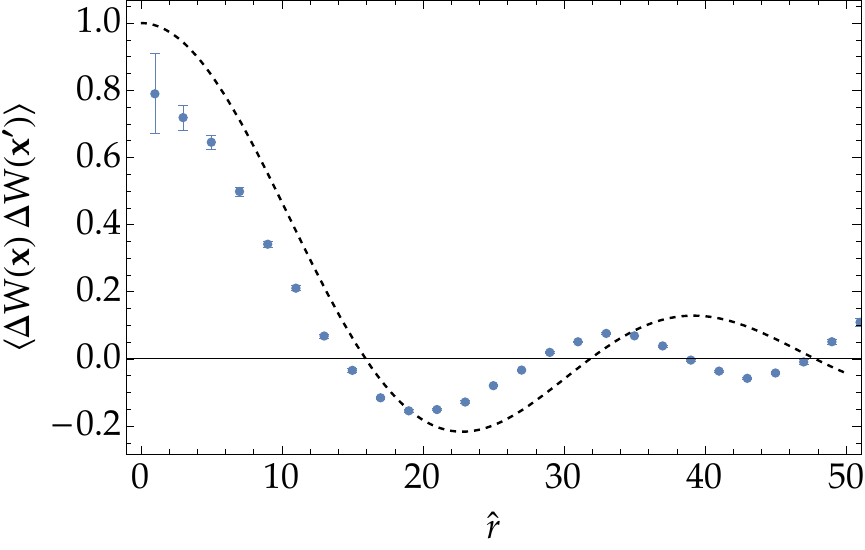}
		\end{minipage}&
		\begin{minipage}{0.3\hsize}
			\centering
			\includegraphics[width=\hsize]{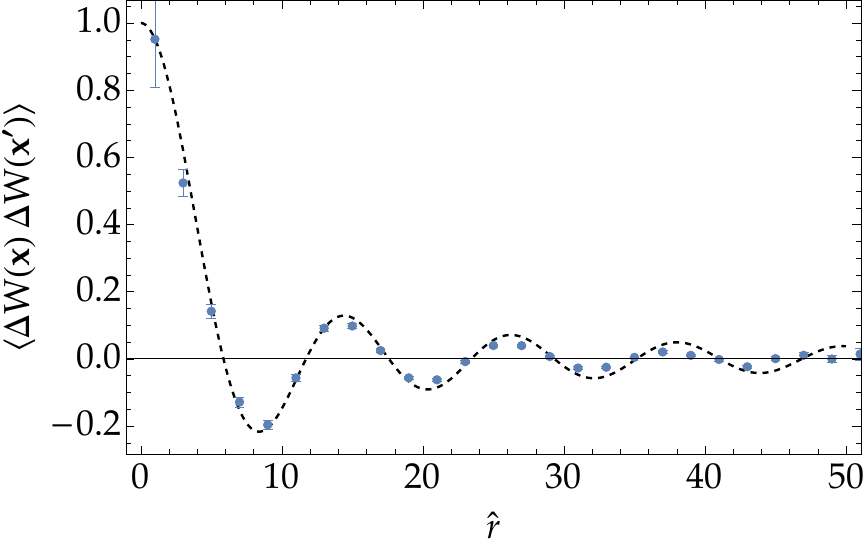}
		\end{minipage}&
    	\begin{minipage}{0.3\hsize}
			\centering
			\includegraphics[width=\hsize]{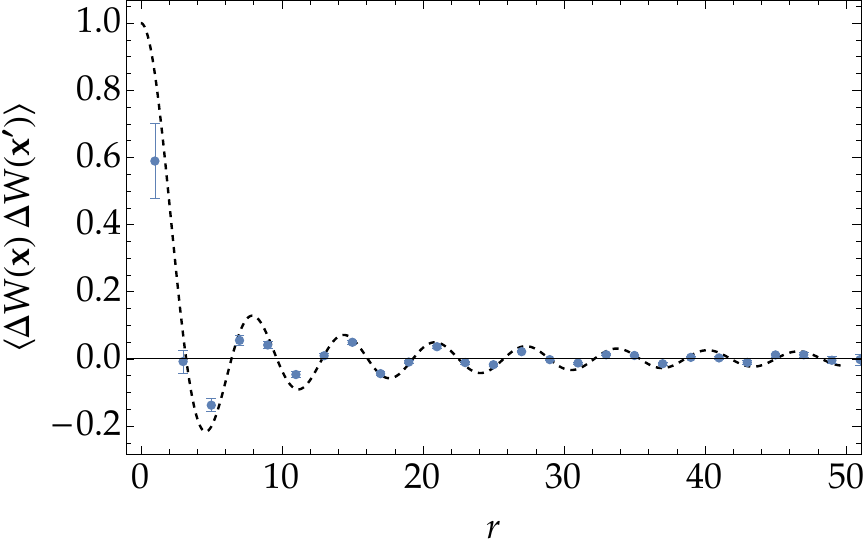}
		\end{minipage}
    \end{tabular}
    \caption{3D plots of the sample of $\Delta W(N,\bfx)$ (top) and the corresponding spatial correlation $\expval{\Delta W(\bfx)\Delta W(\bfx')}$ as a function of $\hat{r}=\frac{N_L}{L}\abs{\bfx-\bfx'}$ (bottom) at $N=3.0$, $4.0$, and $4.6$ corresponding to $n_\sigma\simeq2.0$, $5.5$, and $10$ from left to right.
    In the bottom panels, the blue dots with error bars represent the numerical results, while the black dashed lines are the theoretical requirement $\sinc\pqty{\frac{2\pi n_\sigma}{N_L}\hat{r}}$.
    The numerical correlation is estimated in the Monte Carlo way, that is, $10^6$ pairs of points $(\bfx,\bfx')$ are randomly chosen, sorted to the $\hat{r}$-bin with the width $\Delta\hat{r}=2$, and the average and the standard error are evaluated for each bin.
    One finds that the realised noise reproduces the theoretical correlation well for large enough $n_\sigma$, while it fails for small $n_\sigma$ due to the discreteness of the lattice.}
    \label{fig: samples of noise maps}
\end{figure}

In Fig.~\ref{fig: samples of noise maps}, we showed samples of the $\Delta W(N,\bfx)$ map at $N=3.0$, $4.0$, and $4.6$ corresponding to $n_\sigma\simeq2.0$, $5.5$, and $10$ for $\sigma=0.1$ and the scale factor normalisation $a(N=0)\sfH L=2\pi$ at the beginning of the simulation $N=0$.
The correlation $\expval{\Delta W(N,\bfx)\Delta W(N,\bfx')}$ calculated for these maps as a function of the (normalised) distance $\hat{r}=\frac{N_L}{L}\abs{\bfx-\bfx'}$ is also compared with the theoretical requirement $\sinc (k_\sigma r)=\sinc\pqty{\frac{2\pi n_\sigma}{N_L}\hat{r}}$ where $\sinc x=\frac{\sin x}{x}$ is the sine cardinal function.
Though the small $n_\sigma$ map fails due to the discreteness of the lattice, the maps for large enough $n_\sigma$ successfully reproduce the required correlation.

We also note that, thanks to our choice of the coarse-graining scale in Eq.~\eqref{eq: XIR}, the noise correlation~\eqref{eq: dW correlation} and hence the noise distribution~\eqref{eq: random_point} are independent of the dynamics and determined only by the time coordinate $N$.
Therefore, we can prepare the noise map data in advance of the simulation.

\subsection[Stochastic-$\delta\calN$]{\boldmath Stochastic-$\delta\calN$}

We have all the ingredients to calculate the inflaton fluctuations
(and their conjugate momenta) in the stochastic formalism. However, the inflatons' fluctuations in themselves are not direct observables in the late universe as they decay into daughter particles and disappear after the end of inflation.
That is why cosmologists often calculate the conserved quantity on a superHubble scale: the curvature perturbation $\zeta$.
Thanks to the $\delta N$ formalism~\cite{Starobinsky:1985ibc,Salopek:1990jq,Sasaki:1995aw,Sasaki:1998ug,Wands:2000dp,Lyth:2004gb,Lyth:2005fi}, the curvature perturbation (on a uniform-density slice) $\zeta$ can be obtained as a fluctuation in the e-folding number, $\delta N$, between an initial flat-slicing hypersurface to a final uniform-density hypersurface.
In the stochastic formalism, the e-folding number is also a stochastic variable labelled $\calN$, which is understood as the \emph{first passage time} to the end of inflation in terms of stochastic calculus.
The techniques in stochastic calculus can reveal the statistics of $\calN$ and hence $\zeta$, known as the \emph{stochastic-$\delta\calN$ approach}~\cite{Fujita:2013cna,Fujita:2014tja,Vennin:2015hra} (see, e.g., Refs.~\cite{Kawasaki:2015ppx,Assadullahi:2016gkk,Vennin:2016wnk,Pattison:2017mbe,Ezquiaga:2018gbw,Noorbala:2018zlv,Firouzjahi:2018vet,Noorbala:2019kdd,Kitajima:2019ibn,Prokopec:2019srf,Ezquiaga:2019ftu,Firouzjahi:2020jrj,De:2020hdo,Ando:2020fjm,Figueroa:2020jkf,Pattison:2021oen,Figueroa:2021zah,Tada:2021zzj,Ezquiaga:2022qpw,Ahmadi:2022lsm,Nassiri-Rad:2022azj,Animali:2022otk,Tomberg:2022mkt,Gow:2022jfb,Rigopoulos:2022gso,Briaud:2023eae,Asadi:2023flu,Tomberg:2023kli,Tada:2023fvd,Tokeshi:2023swe,Raatikainen:2023bzk} for its application).

In \ac{STOLAS}, practically simulations proceed on flat slices and will be stopped well before the end of inflation for all grid points.
At that end time, the grid spacing is larger than the Hubble scale.
Therefore, one can understand the final configuration of the simulation as an \emph{initial} flat-slicing hypersurface with superHubble fluctuations for the $\delta N$ scheme.
After that, one solves the \ac{EoM}~\eqref{eq: single noise EoM} until some final uniform-density slice around the end of inflation\footnote{Practically, we choose the end slice at the end of inflation $\epsilon_1=\varpi^2/(2\Mpl^2H^2)=1$ for chaotic inflation and a uniform-inflaton surface for Starobinsky's linear model for simplicity. See Sec.~\ref{sec: Starobinskys linear} for the detailed condition for Starobinsky's linear model.} with the independent noise $\xi(N,\bfx)$ for each grid as they are separated further than the Hubble scale and then obtains $\zeta(\bfx)$ for each point as the time fluctuation $\delta N(\bfx)$.
One may want to understand $\zeta(\bfx)$ as the coarse-grained curvature perturbation over all the smaller scales than the grid spacing.
In that case, letting $\calN(\varphi,\varpi)$ the \emph{stochastic} e-folding number from the initial field values $(\varphi,\varpi)$ to the end hypersurface, the coarse-grained curvature perturbation $\zeta_\cg(\bfx)$ is calculated as
\bae{
    \zeta_\cg(\bfx)=\expval{\calN(\varphi(\bfx),\varpi(\bfx))}-\overline{\expval{\calN(\varphi(\bfx),\varpi(\bfx))}},
}
where $(\varphi(\bfx),\varpi(\bfx))$ are the field values at $\bfx$ at the end of the simulation, the braket stands for the ensemble average, and the overline represents the grid average,\footnote{One may include the volume effect in these averagings. See Ref.~\cite{Animali:2024jiz} for details.} $\overline{X(\bfx)}=\frac{1}{N_L^3}\sum_\bfx X(\bfx)$. 
If one sets the simulation end time so that all grid points are well converged to a slow-roll attractor behaviour as we will do in this paper, the ensemble average can be approximated by the \emph{classical} e-folds $N_\cl$ obtained without noise:
\bae{
    \zeta_\cg(\bfx)\approx N_\cl(\varphi(\bfx),\varpi(\bfx))-\overline{N_\cl(\varphi(\bfx),\varpi(\bfx))}.
}
We adopt \ac{RK4} in the calculation of $N_\cl$. Hereafter, we will omit the subscript $\cg$ for brevity.

\begin{figure}
	\begin{tabular}{c}
		\begin{minipage}{0.5\hsize}
			\centering
			\includegraphics[width=0.95\hsize]{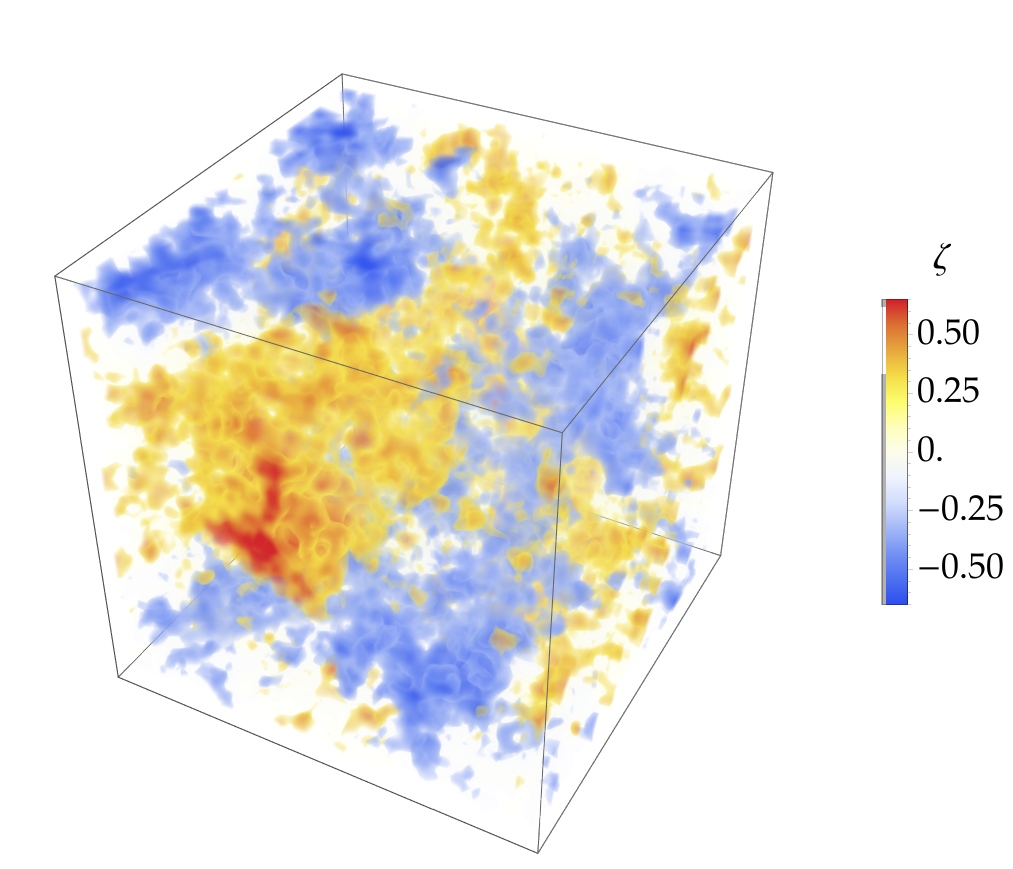}
		\end{minipage}
		\begin{minipage}{0.5\hsize}
			\centering
			\includegraphics[width=0.95\hsize]{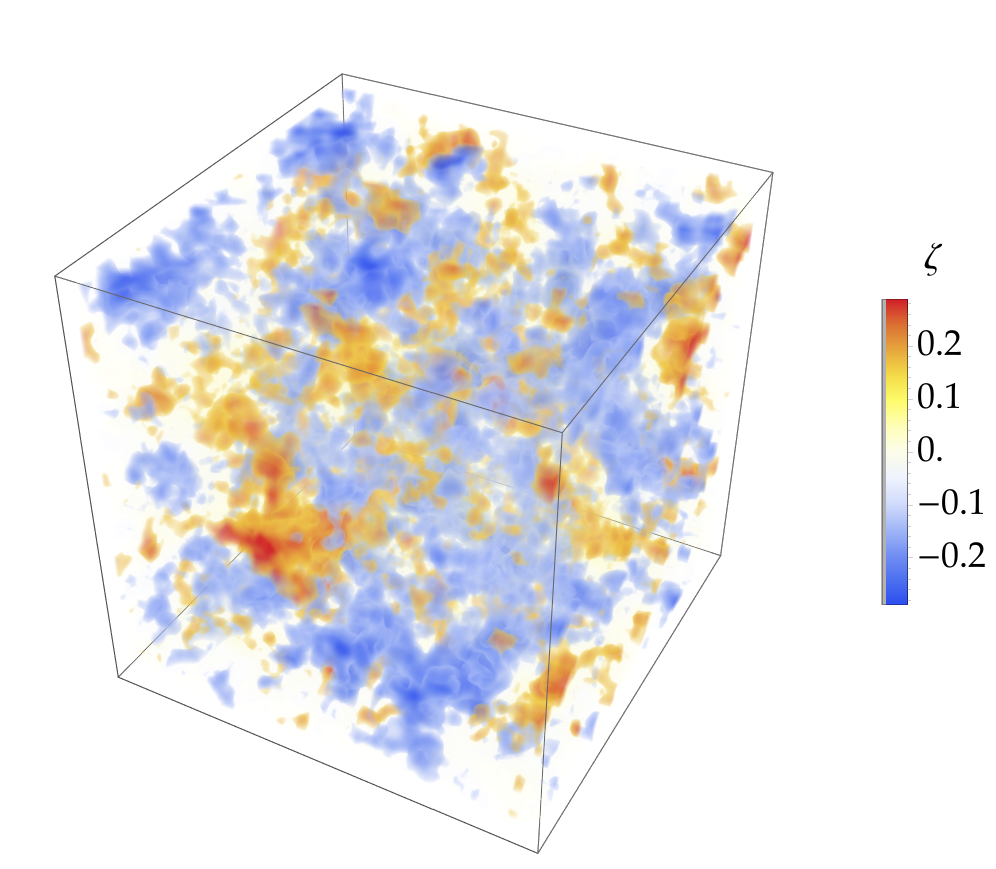}
		\end{minipage}
	\end{tabular}
	\caption{Sample maps of $\zeta(\bfx)$ for the chaotic (left) and Starobinsky's linear (right) models with the parameter sets given in Sec.~\ref{sec: Primary statistics}.}
	\label{fig: zeta chaotic and Starobinsky}
\end{figure}

In Fig.~\ref{fig: zeta chaotic and Starobinsky}, we show sample maps of $\zeta(\bfx)$ for two different models described in the following sections in detail. 
Though we adopt the same noise map in these two models, the resultant $\zeta(\bfx)$ is obviously distinct in both the amplitude and the scale dependence.

\section{Primary statistics}\label{sec: Primary statistics}

In this section, we check the behaviour of \ac{STOLAS} by comparing the primary statistics of $\zeta$, the power spectrum and the skewness of the \ac{PDF}, with the standard perturbation theory.
We consider two models of inflation: chaotic inflation and Starobinsky's linear potential.

\subsection{Calculation of power spectrum}

Let us first discuss how to convert the discrete map $\zeta(\bfx)$ obtained in the lattice simulation (see Fig.~\ref{fig: zeta chaotic and Starobinsky} for example) to the (dimensionless) power spectrum $\calP_\zeta(k)$ defined in the continuous Fourier space as usual.
The discrete Fourier transformation of $\zeta$ is defined similarly to Eq.~\eqref{eq: DFT of dW} as
\bae{
    \zeta_\bfn=\sum_\bfx\zeta(\bfx)\ee^{-i\frac{2\pi}{L}\bfn\cdot\bfx} \qc
    \zeta(\bfx)=\frac{1}{N_L^3}\sum_\bfn\zeta_\bfn\ee^{i\frac{2\pi}{L}\bfn\cdot\bfx}.
}
According to the ergodic theorem, the stochastic average of the one-point variance $\expval{\zeta^2(\bfx)}$ (both in the discrete and the continuous space) can be approximated by the grid-average variance as
\bae{
    \expval{\zeta^2(\bfx)}\approx\overline{\zeta^2(\bfx)}.
}
The grid-average variance is related to the discrete Fourier mode by
\bae{
    \overline{\zeta^2(\bfx)}=\frac{1}{N_L^6}\sum_{\bfn,\bfm}\zeta_\bfn\zeta_\bfm\overline{\ee^{i\frac{2\pi}{L}(\bfn+\bfm)\cdot\bfx}}=\frac{1}{N_L^6}\sum_\bfn\abs{\zeta_\bfn}^2,
}
where we used $\overline{\ee^{i\frac{2\pi}{L}(\bfn-\bfm)\cdot\bfx}}=\delta_{\bfn\bfm}$ and the reality condition $\zeta_{-\bfn}=\zeta_\bfn^*$.
On the other hand, the continuous-space variance can be calculated as the integral of the power spectrum in the logarithmic wavenumber,
\bae{
    \expval{\zeta^2(\bfx)}=\int_{\ln k_\umin}^{\ln k_\umax}\dd{\ln k}\calP_\zeta(k),
}
within the relevant wavenumber range $k\in[k_\umin,k_\umax]$.
The consistency hence leads to the following relation between the continuous-limit power spectrum and the discrete Fourier mode:
\bae{\label{eq: calPz in STOLAS}
    \calP_\zeta\pqty{\frac{2\pi}{L}n}\approx\frac{1}{N_L^6\Delta\ln n}\sum_{\abs{\ln m-\ln n}\leq\frac{\Delta\ln n}{2}}\abs{\zeta_\bfm}^2,
}
where $\Delta\ln n$ is a certain binning parameter.

\subsection{Chaotic inflation}

Let us first see the simplest example: chaotic inflation with the quadratic potential, 
\begin{align}
  V(\phi) = \frac{1}{2} m^2 \phi^2,
\end{align}
where $m$ is the mass of the inflaton.
We chose the mass and the initial condition of the inflaton as $m=2.11\times 10^{-2}\Mpl$ and $(\varphi_\ui,\varpi_\ui)=(11\Mpl, -\sqrt{2/3}m\Mpl)$.
We have chosen these values that correspond to $\sim30$ e-folds before the end of inflation in the classical trajectory because we are interested in small-scale physics in this work.
For the single-field slow-roll models such as chaotic inflation, the inflaton's power spectrum as the noise amplitude is calculated up to the next-to-leading order in the slow-roll approximation as\footnote{One sees that it reduces to the standard noise-amplitude $\calP_\phi^{1/2}\sim H/(2\pi)$ at the leading-order in the slow-roll approximation ($\epv,\hv\to0$). We found that this simple noise amplitude leads to a slightly inconsistent result with the linear perturbation theory. It is necessary to incorporate up to the first order of the slow-roll approximation into the noise amplitude in order for a consistent result within the numerical error.}
\begin{align}\label{eq: calPphi SR}
  {\cal P}_{\phi} = 
  \qty(\frac{H}{2\pi})^2
  \qty(\frac{\sigma \sfH}{2 H})^{- 6 \epv + 2\hv}
  \qty[1 + \epv \qty(10 - 6 \gamma -12\ln{2})
  - 2 \hv\qty(2-\gamma-2\ln{2}) ],
\end{align}
where $\gamma$ is a Euler’s constant (see also Appendix~\ref{sec: calPphi} for detailed derivation).
The slow-roll parameters are given by 
\begin{align}
  \epv = \frac{\Mpl^2}{2} \qty(\frac{V'}{V})^2,
  \quad
  \hv = \Mpl^2 \frac{V''}{V}.
\end{align}
Note that these slow-roll parameters as well as the Hubble parameter should be understood as functions of the field values $\varphi$ and $\varpi$ at the spatial grid and the time step of interest.
$\epsilon_V=\eta_V$ does hold in the quadratic model.

\begin{figure}
	\centering
	\begin{tabular}{c}
  		\begin{minipage}{0.5\hsize}
			\centering
			\includegraphics[width=0.9\hsize]{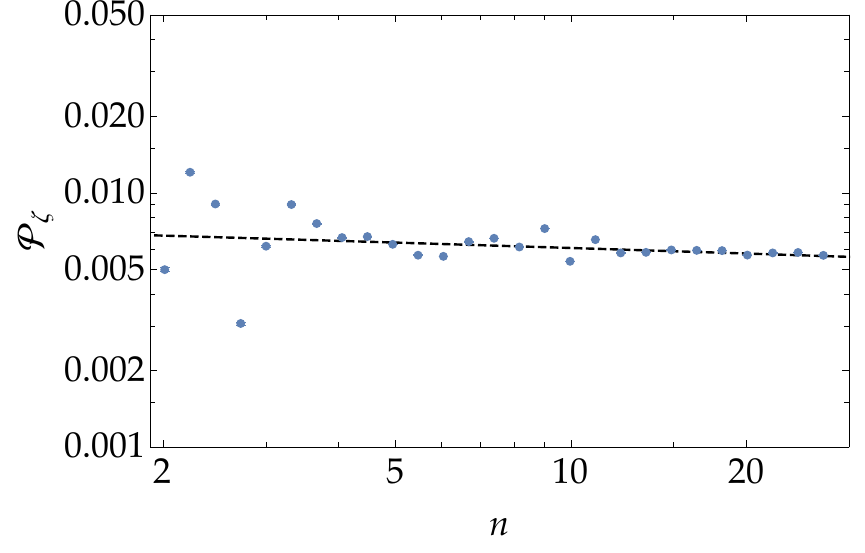}
		\end{minipage}
		\begin{minipage}{0.5\hsize}
			\centering
			\includegraphics[width=0.9\hsize]{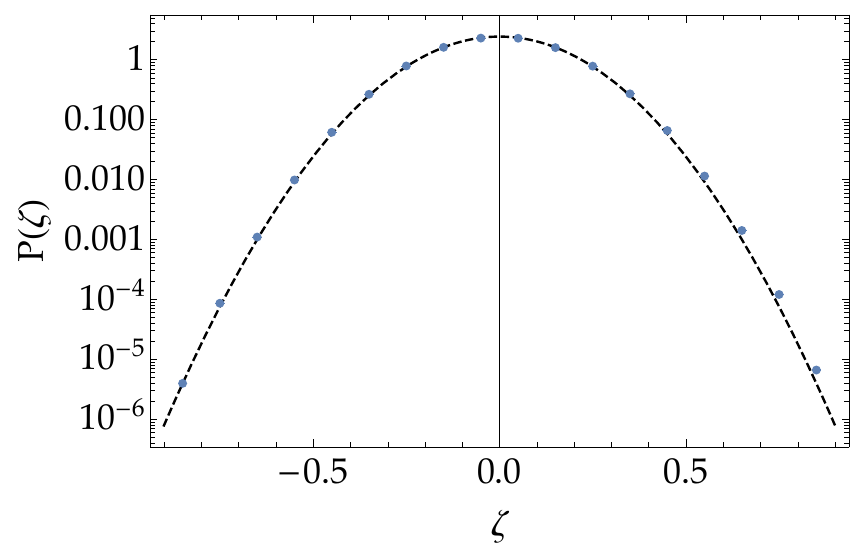}
		\end{minipage}
	\end{tabular}
    \caption{The power spectrum $\calP_\zeta$ (left) and \ac{PDF} $\uP(\zeta)$ (right) of the curvature perturbation in chaotic inflation.
	We set parameters as $\Delta\ln{n}=0.1$, $N_L^3=64^3$, and $\sigma=0.1$.
	The blue points represent the simulation result, while the black dashed line represents the estimation in the linear perturbation theory through the Mukhanov--Sasaki equation in the left panel and the Gaussian fitting with the standard deviation $\sigma_{\zeta}\simeq 0.027$ in the right panel.
	}
    \label{fig:StatChaotic}
\end{figure}

In the left panel of Fig.~\ref{fig:StatChaotic}, we show the power spectrum $\calP_\zeta$~\eqref{eq: calPz in STOLAS} at the end of the inflation.
We simulate with $N_L^3=64^3$ points and the coarse-graining scale $\sigma =0.1$.
The average and the standard error of the power spectrum are estimated with 1000 realisations of the three-dimensional map of curvature perturbation, though the error bars are too small to distinguish.
In this plot, the blue points are our simulation results and the black dashed line is the linear-perturbation estimation through the Mukhanov--Sasaki equation.
As expected from Fig.~\ref{fig: samples of noise maps}, while they are scattered for small wavenumbers, the simulation results are well-consistent with the linear perturbation theory for large-enough wavenumbers $n\gtrsim4$.

We also show the one-point \ac{PDF} of $\zeta(\bfx)$ in the right panel of Fig.~\ref{fig:StatChaotic}. 
The blue dots show our simulation result and the black dashed line is its Gaussian fitting with the standard deviation $\sigma_\zeta=0.027$ calculated by the data.
The result is almost Gaussian but one can detect small non-Gaussianity, e.g., in terms of the non-linearity parameter defined by
\begin{align}\label{eq: fNL}
    \fNL = \frac{5}{18} \frac{\expval{\zeta^3(\bfx)}}{\expval{\zeta^2(\bfx)}^2},
\end{align}
equivalent to the standard non-linearity parameter in the local-type assumption~\cite{Komatsu:2001rj}.
In particular, the so-called Maldacena consistency relation predicts the relation $\fNL=\frac{5}{12}(1-\ns)$ where $\ns-1=\dv*{\ln\calP_\zeta}{\ln k}$~\cite{Maldacena:2002vr}.
We numerically evaluate the Mukhanov--Sasaki equation and read the fitting parameters of the power-law assumption $\calP_\zeta=A_{\zeta} n^{\ns-1}$ as $A_{\zeta}=0.007247$ and $\ns=0.93048$, which means $\fNL=2.89\times 10^{-2}$.
Our simulation shows $\fNL=(3.06\pm 0.43)\times 10^{-2}$ which also shows the consistency with the standard perturbation theory as well as the power spectrum.\footnote{It is known that the reproduction of Maldacena's consistency relation requires the cubic correlation both from the $\delta N$ non-linear mapping and the inflaton's intrinsic non-Gaussianity in the $\delta N$ formalism (see, e.g., Ref.~\cite{Vernizzi:2006ve,Yokoyama:2007uu,Yokoyama:2007dw}). While the $\delta N$ non-linear mapping is taken into account by numerically solving the non-linear \ac{EoM}, the intrinsic non-Gaussianity seems not as we adopt the Gaussian noise. However, the intrinsic non-Gaussianity in the squeezed limit can actually be understood as the background-dependent noise amplitude~\cite{Tada:2016pmk,Abolhasani:2018gyz} and hence we do take its account via the noise amplitude $\calP_\phi$ evaluated at each grid point, which is why we obtained the consistent result with the standard perturbation theory. The equilateral-type non-Gaussianity cannot be included by the Gaussian noise for example, though it is known that that type of non-Gaussianity is negligible in the canonical single-field models (see, e.g., Ref.~\cite{Chen:2010xka}).}

\subsection{Starobinsky's linear potential}\label{sec: Starobinskys linear}

Let us also study Starobinsky's linear-potential model~\cite{Starobinsky:1992ts} as a toy model beyond the slow-roll dynamics.
The potential is given by 
\bae{
    V(\phi)=\bce{
        V_{0} + A_{+} (\phi - \phi_{0})
        &\text{for $\phi > \phi_{0}$},
        \\
        V_{0} + A_{-} (\phi - \phi_{0})
        &\text{for $\phi\leq\phi_{0}$},
    }
}
where $V_{0}$ and $A_{\pm}>0$ are parameters and $\phi_{0}$ is a broken point of potential.
If $A_+>A_-$, the terminal velocity of the inflaton before $\phi_0$ is larger than that after $\phi_0$ and hence the inflaton experiences the friction-dominated phase called \emph{ultra-slow-roll (USR)} right after $\phi_0$.
The \ac{USR} phase can amplify the curvature perturbation and is compatible with the \ac{PBH} scenario.

While the noise power is well approximated by $\calP_\phi^{1/2}=H/(2\pi)$ in the first stage before $\phi_0$, the slow-roll result~\eqref{eq: calPphi SR} is not applicable for the second stage after $\phi_0$ due to the violation of the slow-roll condition.
In the constant Hubble and neglecting noise approximations, the Mukhanov--Sasaki equation has an analytic solution at each phase and can be connected at $\phi_0$, which leads to the analytic expression of $\calP_\phi$ for the second stage as (see also Ref.~\cite{Pi:2022zxs})
\bae{
    &\calP_{\phi,2}(k=\sigma aH) \nonumber \\
    &=\pqty{\frac{H}{2\pi}}^2\times\frac{1}{2 \alpha ^6 \Lambda ^2 \sigma ^6}\left[3 \left(\Lambda ^2 \left(\alpha ^4 (4 \alpha -7) \sigma ^6+\alpha ^3 (7 \alpha -16)
   \sigma ^4+(3-12 \alpha ) \sigma ^2-3\right) \right.\right. \nonumber \\
   &\qquad\quad+\Lambda  \left(2 (5-2 \alpha ) \alpha ^4
   \sigma ^6+2 (14-5 \alpha ) \alpha ^3 \sigma ^4+6 (4 \alpha -1) \sigma ^2+6\right) \nonumber \\
   &\qquad\quad\left.-3
   \left(\alpha ^4 \sigma ^6-(\alpha -4) \alpha ^3 \sigma ^4+(4 \alpha -1) \sigma
   ^2+1\right)\right) \cos (2 (\alpha -1) \sigma ) \nonumber \\
   &\qquad+\left(\sigma ^2+1\right) \left(-18
   \Lambda  \left(\alpha ^2 \sigma ^2+1\right)^2+9 \left(\alpha ^2 \sigma
   ^2+1\right)^2+\Lambda ^2 \left(2 \alpha ^6 \sigma ^6+9 \alpha ^4 \sigma ^4+18 \alpha ^2
   \sigma ^2+9\right)\right) \nonumber \\
   &\qquad+6 \sigma  \left(\alpha ^5 (\Lambda -1) \Lambda  \sigma ^4
   \left(\sigma ^2-1\right)+\alpha ^4 \left(7 \Lambda ^2-10 \Lambda +3\right) \sigma
   ^4-\alpha ^3 \left(4 \Lambda ^2-7 \Lambda +3\right) \sigma ^2 \left(\sigma ^2-1\right) \right. \nonumber \\
   &\qquad\quad\left.\left.-3
   \alpha  (\Lambda -1)^2 \left(\sigma ^2-1\right)-3 (\Lambda -1)^2\right) \sin (2 \sigma -2
   \alpha  \sigma )\right],
}
where $\Lambda\coloneqq A_+/A_-$, $H\simeq\sfH\simeq H_0\coloneqq\sqrt{V_0/(3\Mpl^2)}$, and $\alpha=\exp(N-\calN_0)$ where $\calN_0$ is the transition time: $\varphi(\calN_0)=\phi_0$.

\begin{figure}
	\centering
	\begin{tabular}{c}
        \begin{minipage}{0.5\hsize}
            \centering
            \includegraphics[width=0.95\hsize]{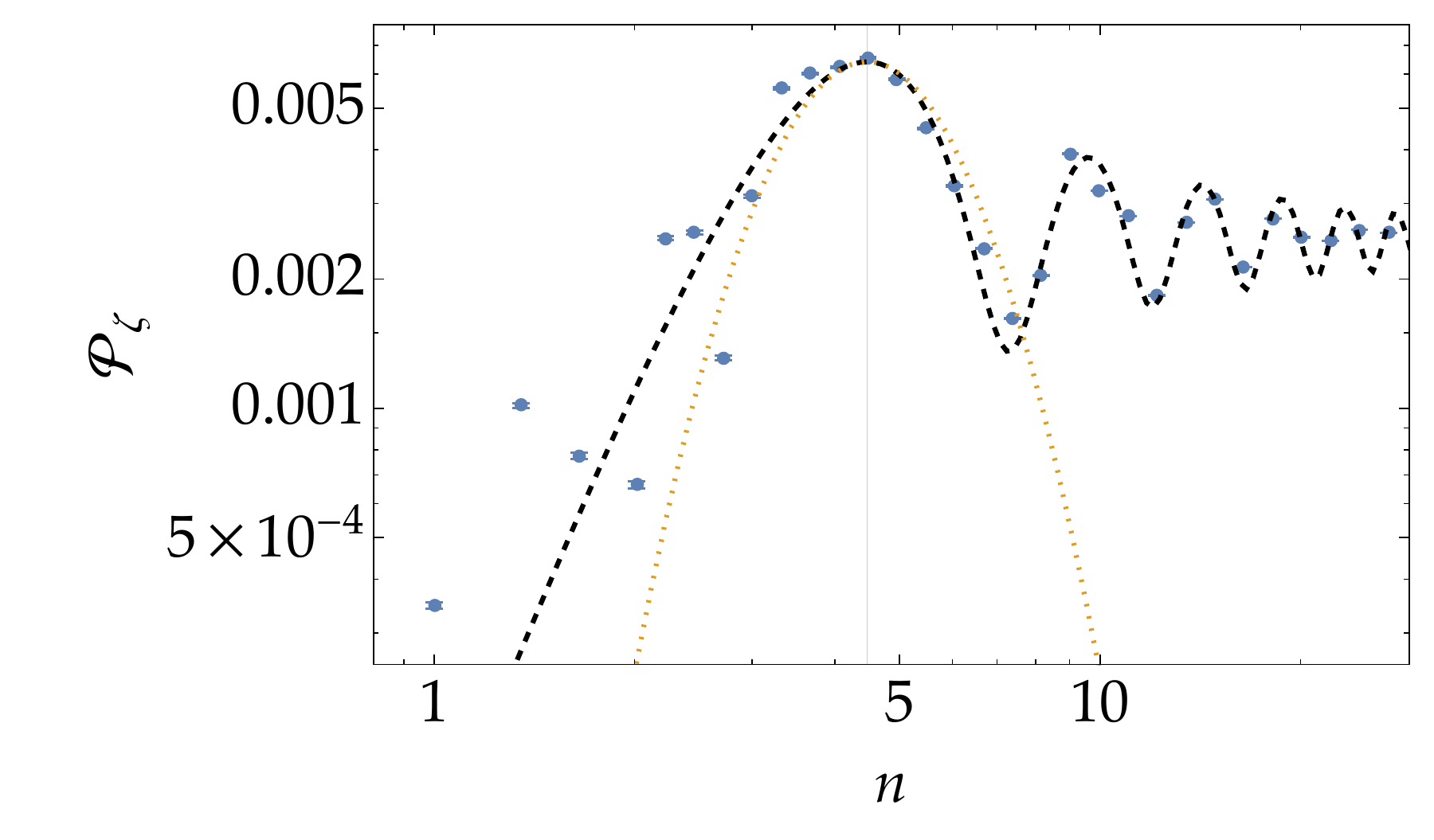}
        \end{minipage}
        \begin{minipage}{0.5\hsize}
            \centering
            \includegraphics[width=0.95\hsize]{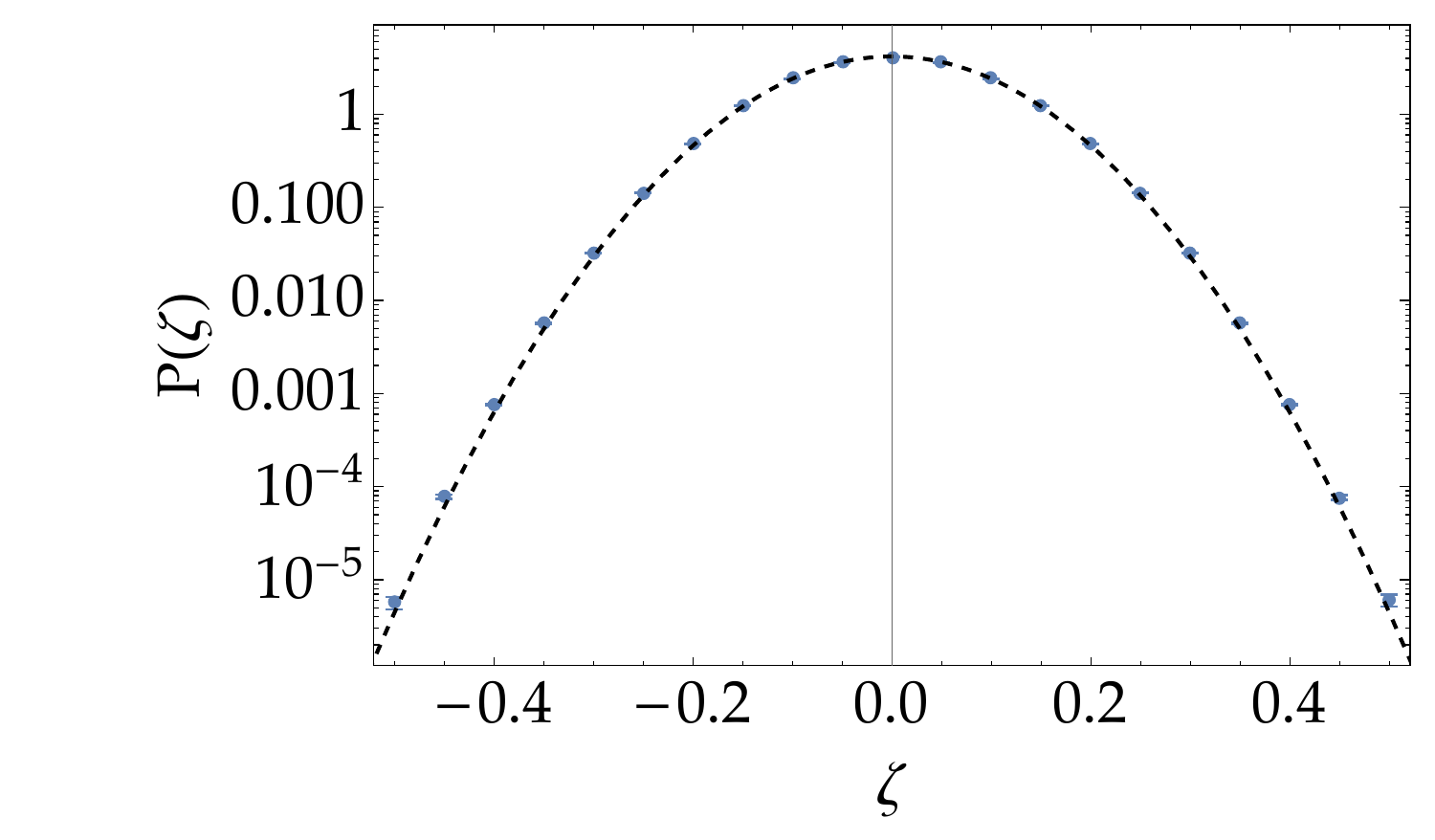}
        \end{minipage}
    \end{tabular}
    \caption{The same plots as Fig.~\ref{fig:StatChaotic} for the Starobinsky's linear model.
    The black dashed line represents the analytical formula~\eqref{eq:PowerStarobinsky} in the linear perturbation theory in the left panel, while it exhibits a Gaussian fitting with the standard deviation $\sigma_\zeta\simeq0.095$ in the right panel.
    In the left panel, the black vertical thin line indicates the peak position of the power spectrum, $n_*\simeq4.47$, and the orange dotted line is a log-normal fitting of the power spectrum about that peak, which are relevant for \ac{PBH} formation in the next section.
	}
    \label{fig: StatStarobinsky}
\end{figure}

In the simulation, we chose the parameters as $H_0=10^{-5}M_\mathrm{Pl}$, $A_+=\sqrt{\frac{9H_0^6}{4\pi^2\times8.5\times10^{-10}}}$, and $A_-=A_+/1700$ and the initial condition as $(\varphi_\ui,\varpi_\ui)=(1.93\times 10^{-2}\Mpl,-5.45\times 10^{-7}\Mpl^2)$ and set the end surface of the $\delta N$ formalism at $\phi_\uth=-1.87\times 10^{-2}\Mpl$, where the inflaton sufficiently asymptotes to the slow-roll attractor behaviour, supposing that the curvature perturbation gets frozen well regardless of whether inflation ends there or not.
In the left panel of Fig.~\ref{fig: StatStarobinsky}, we show the power spectrum with the same lattice setup and plot style as chaotic inflation.
For the UV dynamics, we adopt the analytic solution of the Mukhanov--Sasaki equation given by
\begin{align}
  \frac{{\cal P}_{\zeta} (k)}{{\cal P}_{\rm IR}} =
  \frac{9 (\Lambda -1)^2}{k^6} (\sin k - k \cos k)^4
  + \left[\frac{3 (\Lambda -1)}{2 k^3}
  \left[\left(k^2-1\right) \sin (2 k)+2 k \cos (2 k)\right]
  + \Lambda \right]^2,
  \label{eq:PowerStarobinsky}
\end{align}
where our parameter choice corresponds to $\Lambda=1700$ and $\mathcal{P}_\mathrm{IR}=8.5\times10^{-10}$.
One again sees their sufficient consistency for $n\gtrsim4$.
In particular, \ac{STOLAS} works well for the peak value at $n_*\simeq4.47$ (indicated by the vertical thin line) relevant for the \ac{PBH} abundance.

In the right panel of Fig.~\ref{fig: StatStarobinsky}, we show the \ac{PDF} of curvature perturbation with a Gaussian fitting (black dashed) for the standard deviation $\sigma_{\zeta}=0.095$.
The simulation result is consistent with the Gaussian distribution as $\fNL=(-1.44\pm2.77)\times10^{-3}$.\footnote{We do not show the $\fNL$ value obtained in the standard perturbation theory because both the power and bispectrum highly depend on the scale and the standard non-linearity parameter is not given by the simple formula~\eqref{eq: fNL}. The value obtained in the simulation should be understood as a mere reference of the Gaussianity of the curvature perturbation.}
We comment on the so-called exponential-tail feature which has recently attracted attention.
For the flat-inflection potential including an \ac{USR} phase, it is suggested that the probability of large curvature perturbations decays only exponentially rather than the Gaussian decay (see, e.g., Refs.~\cite{Pattison:2017mbe,Ezquiaga:2019ftu,Figueroa:2020jkf}).
The non-Gaussian feature represented by this exponential tail in the \ac{USR} model is sensitive to the end of the \ac{USR} phase~\cite{Cai:2018dkf,Passaglia:2018ixg}.
In our setup, the \ac{USR} phase smoothly connects with the second slow-roll phase. 
Such a smooth transition is known to erase the non-Gaussian feature. This is why the simulation result is almost Gaussian.
One may add a second breaking point in the potential to force the \ac{USR} phase to end at that point, and then the exponential-tail feature is expected.

\section{Importance sampling for primordial black holes}\label{sec: Importance sampling}

In this section, we further show the power of \ac{STOLAS}, simulating the formation of \acp{PBH}. \acused{BH} 
The \acp{PBH} are hypothetical black holes produced by overdense regions in the radiation-dominated era through gravitational collapse, proposed by Hawking and Carr~\cite{Hawking:1971ei,Carr:1974nx,Carr:1975qj} (see also Refs.~\cite{Carr:2020gox,Escriva:2022duf,Yoo:2022mzl,Carr:2023tpt} for several reviews).
Recently, they have had much motivation from both the theoretical and observational sides.
The most famous motivation is the candidate of \ac{DM}. The current observational constraints on \acp{PBH} show the window for the whole \ac{DM} in the mass range $\sim10^{20}\,\si{g}$, which corresponds to the subHertz scalar-induced \ac{GW} as an interesting target of the space-based future \ac{GW} telescopes such as LISA~\cite{amaro2017laser}, Taiji~\cite{Ruan:2018tsw}, TianQin~\cite{TianQin:2015yph}, and DECIGO~\cite{Kawamura:2011zz}.
As other motivations, they can explain the seed of supermassive black holes~\cite{Carr:2018rid}, the microlensing events observed by OGLE~\cite{Niikura:2019kqi}, the hypothetical ninth planet in the solar system~\cite{PhysRevLett.125.051103, Witten:2020ifl}, the origin of faint supernovae which are the calcium-rich gap transients~\cite{Smirnov:2022zip}, etc.
Subsolar compact objects recently suggested by the observations of the merger \acp{GW} are also the interesting possibility of \acp{PBH} because the astrophysical \acp{BH} cannot be lighter than the solar mass~\cite{Phukon:2021cus,LIGOScientific:2022hai,Prunier:2023cyv}.

It is suggested that the \ac{PBH} formation is characterised not only by the absolute value of the curvature perturbation but also its spatial profile (see, e.g., Refs.~\cite{Atal:2019erb,Escriva:2019phb}).
\ac{STOLAS} can get the spatial map of $\zeta$ and hence sample the \ac{PBH} formation in principle. 
However, the direct sampling of \acp{PBH} is non-realistic because \acp{PBH} should be extremely rare ($\sim10\sigma$ rarity, roughly speaking) at their formation time not to overclose the universe. In this paper, we introduce the importance sampling technique (see, e.g., Ref.~\cite{Jackson:2022unc}) to \ac{STOLAS} to efficiently sample the rare objects.

In this section, we first review the \ac{PBH} formation in the so-called peak theory. 
We then introduce the importance sampling method in the context of \ac{STOLAS} and show some results in the \ac{PBH} abundance.

\subsection{Review of PBH formation}

In the radiation-dominated era, basically the subHubble density contrast cannot gravitationally grow due to the radiation pressure.
However, if the density contrast is as large as $\delta_\uth\sim p/\rho=1/3$ at the horizon reentry, the gravitational force can overcome the pressure and the overdensity collapses to a \ac{BH} right after the horizon reentry.
The formed \ac{BH} is called \emph{\acf{PBH}}.\footnote{Several other mechanisms to produce \acp{PBH} have been proposed so far: false-vacuum bubbles~\cite{Hawking:1982ga,Kodama:1982sf,Moss:1994pi,Khlopov:1998nm,Khlopov:1999ys,Deng:2017uwc,Kusenko:2020pcg,Kitajima:2020kig,Maeso:2021xvl}, string-wall networks~\cite{Ferrer:2018uiu,Gelmini:2022nim,Gelmini:2023ngs,Gouttenoire:2023gbn,Dunsky:2024zdo}, the collapse of isocurvature fluctuations~\cite{Passaglia:2021jla,Yoo:2021fxs}, the quark confinment~\cite{Dvali:2021byy}, etc.} Its mass is roughly given by the horizon mass $M_H=\frac{4\pi}{3}\rho H^{-3}$, which is calculated as (see, e.g., Ref.~\cite{Tada:2019amh})
\bae{\label{eq: Mk}
	M_H(k)=10^{20}\,\si{g}\times\pqty{\frac{g_*}{106.75}}^{-1/6}\pqty{\frac{k}{\SI{1.56e13}{Mpc^{-1}}}}^{-2},
}
for the horizon reentry of the comoving mode $k$: $aH/k=1$.
Here, $g_*$ is the effective \acp{DoF} for energy density at the horizon reentry and we suppose that it is approximately equivalent to that for entropy.

Since Carr's simplest criterion $\delta_\uth\sim1/3$, the detailed condition of the \ac{PBH} formation has been investigated both theoretically and numerically. There, the \emph{compaction function}~\cite{PhysRevD.60.084002, PhysRevD.91.084057}, the difference between the Misner--Sharp mass $M_\MS$ and the expected mass $M_\uF$ in the background universe, or equivalently, the volume average of the density contrast at its horizon reentry, 
plays a key role.
In the spherically symmetric case, it is given by\footnote{In our notation, the positive $\zeta$ corresponds to the overdensity.}
\bae{\label{eq: compaction}
    \calC(r)=2G\frac{M_\MS-M_\uF}{R}=\frac{2}{3}\qty[1-\qty(1+r\zeta^\prime(r))^2],
}
where $r$ is the comoving radius from the spherical centre and $R(r)=a\ee^{\zeta(r)}r$ is the corresponding areal radius.
According to the latest works~\cite{Atal:2019erb, Escriva:2019phb}, the average compaction function defined by
\bae{\label{eq: averagecompaction}
    \bar{\calC}_\um=\frac{\displaystyle 4\pi\int^{R(r_\um)}_0 \calC(r)\Tilde{R}^2(r)\dd{\Tilde{R}(r)}}{\displaystyle\frac{4\pi}{3}R^3(r_\um)}
}
gives a relatively profile-independent criterion, $\bar{\calC}_\uth=2/5$. Here,
$r_\um$ is the innermost maximum radius of the compaction function. 

The numerical simulations in general relativity suggest that the \ac{PBH} mass follows the scaling behaviour~\cite{Choptuik:1992jv,Evans:1994pj,Koike:1995jm,Niemeyer:1997mt,Niemeyer:1999ak,Hawke:2002rf,Musco:2008hv},
\bae{\label{eq: M_PBH}
	M_\PBH\simeq K\left(\bar{\calC}_\um-\bar{\mathcal{C}}_\uth\right)^{\gamma}M_H,
}
where $K$ is an order-unity factor, $\gamma\simeq0.36$ is the universal power index, and $M_H$ is the horizon mass at the horizon reentry $R(r_\um)H=1$. We adopt $K=1$ in this work for simplicity.
It is practically useful to rewrite the horizon mass~\eqref{eq: Mk} in terms of the box size $L$ and the renormalised maximal areal radius $\tilde{R}_\um=\frac{N_L}{aL}R(r_\um)=\tilde{r}_\um\ee^{\zeta(r_\um)}$ as
\bae{\label{eq: M_Horizon}
	M_H(k=r_\um^{-1}\ee^{-\zeta(r_\um)})=\SI{2.4e+22}{g}\times\pqty{\frac{g_*}{106.75}}^{-1/6}\pqty{\frac{L}{10^{-12}\,\si{Mpc}}}^2\pqty{\frac{\tilde{R}_\um}{N_L}}^2.
}

The current energy density ratio $f_\PBH$ of \acp{PBH} to total \ac{DM} in each logarithmic mass bin is defined by
\bae{\label{eq: fPBH}   
    f_\PBH(M)\dd{\ln M}=\frac{M n_\PBH(M)}{3\Mpl^2 H_0^2\Omega_\DM}\dd{\ln M},
}
where $\Omega_\DM$ is the \ac{DM} density parameter, $n_\PBH$ is the comoving number density of \acp{PBH}, and $H_0$ is the current Hublle parameter.
Below, we calculate the probability $p_\ut(M)\dd{\ln M}$ with which a single \ac{PBH} within the mass range $\ln M_\PBH\in\qty[\ln M-\frac{1}{2}\dd{\ln M},\ln M+\frac{1}{2}\dd{\ln M}]$ is realised in the simulation box $L^3$, i.e., $n_\PBH(M)=p_\ut(M)/L^3$.
One then finds a numerical formula for the \ac{PBH} abundance:
\bae{\label{eq: fPBH in importance sampling}
	f_\PBH(M)=\pqty{\frac{\Omega_\DM h^2}{0.12}}^{-1}\pqty{\frac{M}{10^{20}\,\si{g}}}\pqty{\frac{L}{10^{-12}\,\si{Mpc}}}^{-3}\pqty{\frac{p_\ut(M)}{\num{6.62e-13}}},
}
where $h=H_0/\pqty{\SI{100}{km.s^{-1}.Mpc^{-1}}}$ is the normalised Hubble parameter.

Let us mention the analytic estimation of the \ac{PBH} abundance so far.
Recently, the so-called peak theory~\cite{Bardeen:1985tr} has attracted attention as the state-of-art scheme for the \ac{PBH} abundance estimation~\cite{Yoo:2018kvb,Yoo:2019pma,Yoo:2020dkz,Kitajima:2021fpq}.\footnote{See also, e.g., Refs.~\cite{Germani:2019zez, Germani:2023ojx} for other recent statistical approaches to the \ac{PBH} abundance.} 
If the curvature perturbation is well approximated by a Gaussian random field, the probability of any configuration of $\zeta(\bfx)$ can be determined only by its power spectrum in principle.
In particular, if the power spectrum is monochromatic as
\bae{\label{eq: monochromatic power}
	\calP_\zeta(k)=A_\us\delta(\ln k-\ln k_*),
}
with the amplitude $A_\us$ and the characteristic scale $k_*$, the estimation becomes much simpler.
For example, the rare peak is typically fit by the spherically symmetric profile,
\bae{\label{eq: typical profile}
	\hat{\zeta}(r)=\tilde{\mu}_2\sinc(k_*r),
}
with the Gaussian random amplitude $\mu_2$.
The expected \ac{PBH} number density reads\footnote{Note that the expression in the original work~\cite{Yoo:2018kvb} suffers from the extra factor of $27$, which propagated to several literatures. It is fixed in our expression.}
\bae{
	n_\PBH^\peak(M)\dd{\ln M}=\frac{k_*^3}{(6\pi)^{3/2}}f\pqty{\frac{\tilde{\mu}_2}{\sqrt{A_\us}}}\uP_\uG(\tilde{\mu}_2,A_\us)\abs{\dv{\ln M}{\tilde{\mu}_2}}^{-1}\dd{\ln M},
}
with the function $f(\xi)$,
\bme{
    f(\xi)=\frac{1}{2}\xi(\xi^2-3)\pqty{\erf\bqty{\frac{1}{2}\sqrt{\frac{5}{2}}\xi}+\erf\bqty{\sqrt{\frac{5}{2}}\xi}} \\
    +\sqrt{\frac{2}{5\pi}}\Bqty{\pqty{\frac{8}{5}+\frac{31}{4}\xi^2}\exp\bqty{-\frac{5}{8}\xi^2}+\pqty{-\frac{8}{5}+\frac{1}{2}\xi^2}\exp\bqty{-\frac{5}{2}\xi^2}}, \label{eq: f}
}
the Gaussian \ac{PDF} $\uP_\uG(x,\sigma^2)=\frac{1}{\sqrt{2\pi\sigma^2}}\ee^{-x^2/(2\sigma^2)}$, and the $\tilde{\mu}_2$-dependence of the \ac{PBH} mass numerically found as\footnote{Note that the scaling behaviour of the \ac{PBH} mass with respect to the amplitude $\tilde{\mu}_2$ is used to derive this relation in Ref.~\cite{Kitajima:2021fpq}, which results in a mere difference of the prefactor $K$.}
\bae{
	\abs{\dv{\ln M}{\tilde{\mu}_2}}=\abs{2\dv{\hat{\zeta}(r_\um)}{\tilde{\mu}_2}+\frac{\gamma}{\tilde{\mu}_2-\tilde{\mu}_{2,\uth}}}\simeq0.282+\frac{0.36}{\tilde{\mu}_2-\tilde{\mu}_{2,\uth}},
}
where $\tilde{\mu}_{2,\uth}\simeq0.615$ is the \ac{PBH} formation threshold for the profile~\eqref{eq: typical profile}.
The \ac{PBH} abundance is hence estimated as
\bme{
    f_\PBH^\peak(M)=\pqty{\frac{\Omega_\DM h^2}{0.12}}^{-1}\pqty{\frac{M}{10^{20}\,\mathrm{g}}}\pqty{\frac{k_*}{1.56\times10^{13}\,\mathrm{Mpc}^{-1}}}^3 \\
    \times\pqty{\frac{\abs{\dv{\ln M}{\tilde{\mu}_2}}^{-1}f\pqty{\frac{\tilde{\mu}_2(M)}{\sqrt{A_\us}}}\mathrm{P}_\mathrm{G}\pqty{\tilde{\mu}_2(M),A_\us}}{\num{1.4e-14}}}. \label{eq: fNL Gauss}
}
Below, we compare the result in our \ac{STOLAS} with this analytic formula.

\subsection{Importance sampling}

The direct sampling of \acp{PBH} is non-realistic as they should be extremely rare.
We here introduce the importance sampling technique to \ac{STOLAS} (see Ref.~\cite{Jackson:2022unc} for its application to the stochastic-$\delta\calN$ formalism).
The simulation of the stochastic process is realised by numerically generating the stochastic noise.
Given the probability distribution of the noise, one can conversely introduce \emph{intentionally} large noise whose true probability is calculable as well.
In this way, one can sample overdensities much more efficiently and correct the obtained histogram to follow the true probability distribution.

In this work, we introduce an offset $\calB(N,\bfx)\Delta N$ called \emph{bias} to the stochastic noise $\Delta W(N,\bfx)$ as
\beae{\label{eq: EoM_bias}    
	&\varphi(N+\Delta N,\bfx)-\varphi(N,\bfx)=\frac{\varpi(N,\bfx)}{H(N,\bfx)}\Delta N+\calP_\phi^{1/2}\pqty{N,k_\sigma(N)}\pqty{\mathcal{B}(N,\bfx)\Delta N+\Delta W(N,\bfx)}, \\
    &\varpi(N+\Delta N,\bfx)-\varpi(N,\bfx)=\pqty{-3\varpi(N,\bfx)-\frac{V'\pqty{\varphi(N,\bfx)}}{H(N,\bfx)}}\Delta N.
}
Respecting the typical profile~\eqref{eq: typical profile}, we adopt the $\sinc$ spatial dependence around the origin at about the time of interest $N_\ub$ as
\bae{\label{eq: bias}
	&\calB\pqty{{N},\bfx}=B(N)S(N,\bfx) \qc
	\bce{
		B(N)=\frac{b}{\sqrt{2\pi}\Delta N_\ub}\exp(-\frac{\pqty{N-N_\ub}^2}{2\pqty{\Delta N_\ub}^2}), \\
		S(N,\bfx)=\sinc(k_\sigma(N)x),
	}
}
where $b$ is the overall amplitude parameter and $\Delta N_\ub$ controlls the duration of the bias around $N_\ub$.
According to the discussion in Sec.~\ref{sec: Discretisation}, the $\sinc$ function can be represented by the superposition of the Fourier modes in the shell $\varsigma(N)$ as
\bae{
	S(N,\bfx)\simeq\frac{1}{\abs{\varsigma(N)}}\sum_{\bfn\in\varsigma(N)}\ee^{i\frac{2\pi}{L}\bfn\cdot\bfx}.
}
Hence the biased \ac{EoM}~\eqref{eq: EoM_bias} is equivalent to replacing the Fourier-space noise as $\Delta W_\bfn(N)\to\Delta W_\bfn(N)+\frac{N_L^3}{\abs{\varsigma(N)}}B(N)\Delta N$. Appropriately choosing the amplitude $b$, maps realising \acp{PBH} around the origin can be sampled more efficiently.

Though the biased \ac{EoM} is different from the original one, the realised samples by this equation denoted $\omega\in\Omega$ in the sample space $\Omega$ can be the solutions of the original one in principle if the whole shifted noise $\Delta W_\bfn(N)+\frac{N_L^3}{\abs{\varsigma(N)}}B(N)\Delta N$ is realised by the original distribution~\eqref{eq: random_point}.
Only its realisation probability is different: $\omega$ is realised with the \emph{sampling probability} $p_\us(\omega)$ by the biased \ac{EoM}, while the true \emph{target probability} $p_\ut(\omega)$ is associated with the original \ac{EoM}.
As $p_\us(\omega)$ can be estimated by the numerical histogram, the target probability $p_\ut(\omega)$ can be obtained by calculating the \emph{weight function} defined by
\bae{
	\calW(\omega)=\frac{p_\ut(\omega)}{p_\us(\omega)}.
}
The sampling probability of $\omega$ is the realisation probability of the noise $\Delta W_\bfn$ as $p_\us(\omega)=p\pqty{\Bqty{\Delta W_\bfn(N)}\mid\omega}$, while $\omega$ is realised by the shifted noise in the true system, i.e., $p_\ut(\omega)=p\left(\Bqty{\Delta W_\bfn(N)+\frac{N_L^3}{\abs{\varsigma(N)}}B(N)\Delta N}\relmiddle{|}\omega\right)$.
Recalling the noise distribution~\eqref{eq: random_point}, the (log of) weight function is hence calculated as
\bae{
	\ln\calW(\omega)=\sum_N\ln w(N),
}
where
\bae{\label{eq: lnwN}
	\ln w(N)&\bmte{=-\sum_{\bfn\in\scrC\cap\varsigma(N)}\frac{\abs{\Delta W_\bfn(N)+\frac{N_L^3}{\abs{\varsigma(N)}}B(N)\Delta N}^2-\abs{\Delta W_\bfn(N)}^2}{\frac{N_L^6}{\abs{\varsigma(N)}}\Delta N} \\
	-\sum_{\bfn\in\scrR\cap\varsigma(N)}\frac{\pqty{\Delta W_\bfn(N)+\frac{N_L^3}{\abs{\varsigma(N)}}B(N)\Delta N}^2-\pqty{\Delta W_\bfn(N)}^2}{2\frac{N_L^6}{\abs{\varsigma(N)}}\Delta N}} \nonumber \\
	&=-\frac{B(N)}{N_L^3}\sum_\bfn\Delta W_\bfn(N)-\frac{1}{2}B^2(N)\Delta N \nonumber \\
	&=-B(N)\Delta W(N,\bfx=\bfzero)-\frac{1}{2}B^2(N)\Delta N.
}
Here we used $2\abs{\scrC\cap\varsigma}+\abs{\scrR\cap\varsigma}=\abs{\varsigma}$, $\Delta W_\bfn=0$ for $\bfn\notin\varsigma$, and $\sum_{\bfn}\Re[\Delta W_\bfn]=\sum_{\bfn}\Delta W_\bfn$.
This is a Gaussian variable of the average and variance given by
\beae{
	&\expval{\ln\calW}=-\frac{1}{2}\sum_NB^2(N)\Delta N\simeq-\frac{1}{2}\int B^2(N)\dd{N}\simeq-\frac{b^2}{4\sqrt{\pi}\Delta N_\ub}, \\
	&\expval{(\ln\calW-\expval{\ln\calW})^2}=\sum_NB^2(N)\Delta N\simeq\int B^2(N)\dd{N}\simeq\frac{b^2}{2\sqrt{\pi}\Delta N_\ub}.
}

Dividing the samples into bins as $\Omega=\bigcup_i\Omega_i$ with respect to quantities of interest, the sampling probability is evaluated by the histogram as
\bae{\label{eq: ps}
	p_\us(\Omega_i)\approx\frac{\abs{\Omega_i}}{\abs{\Omega}}.
}
The target probability is then obtained by multiplying it by the average weight in the corresponding bin:
\bae{
	p_\ut(\Omega_i)=\expval{\calW(\omega\in\Omega_i)}p_\us(\Omega_i).
}
Practically, the average weight is difficult to be estimated by the direct sample average as the weight varies by many orders of magnitude.
Rather the log of the weight can be expected to well follow the Gaussian distribution in each bin like as the unbinned one~\eqref{eq: lnwN}.
Supposing its lognormal distribution, the average weight can be estimated more accurately by the log average and variance within the propagated error as (see Ref.~\cite{Jackson:2022unc})
\bae{\label{eq: lognormal estimator}
	\expval{\calW(\omega\in\Omega_i)}=\exp\pqty{\overline{\ln\calW_i}+\frac{\sigma_{\ln\calW_i}^2}{2}}\abs{\exp[\pm\sqrt{\frac{\sigma_{\ln\calW_i}^2}{\abs{\Omega_i}}+\frac{\sigma_{\ln\calW_i}^4}{2\abs{\Omega_i}-2}}]-1},
}
where
\bae{
	\overline{\ln\calW_i}=\frac{1}{\abs{\Omega_i}}\sum_{\omega\in\Omega_i}\ln\calW(\omega) \qc
	\sigma_{\ln\calW_i}^2=\frac{1}{\abs{\Omega_i}-1}\sum_{\omega\in\Omega_i}(\ln\calW(\omega)-\overline{\ln\calW_i})^2.
}

We stress that the choice of the bias form~\eqref{eq: typical profile} does not change the final result such as the \ac{PBH} abundance but only affects the sampling efficiency.
A wrong choice of the bias merely reduces the efficiency and enlarges the sampling errors but the infinite sampling anyway leads to the true statistics.
The validity of the bias choice can be indicated by the correlation between the weight value and the quantity of interest such as the left panel of Fig.~\ref{fig: density histogram}.

\subsection{Results and Discussion}

\begin{figure}
	\centering
    \begin{tabular}{cc}
        \begin{minipage}{0.48\hsize}
            \includegraphics[width=0.95\hsize]{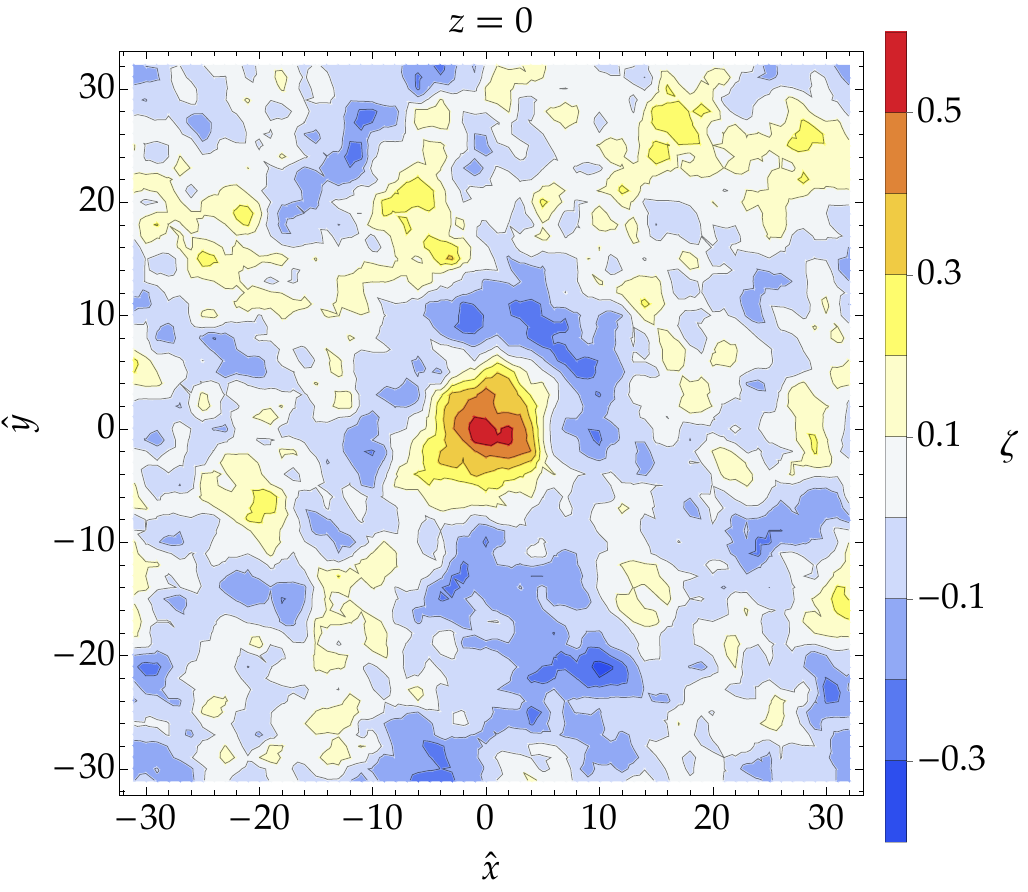}
        \end{minipage}&
        \begin{minipage}{0.48\hsize}
            \centering
        	\includegraphics[width=0.95\hsize]{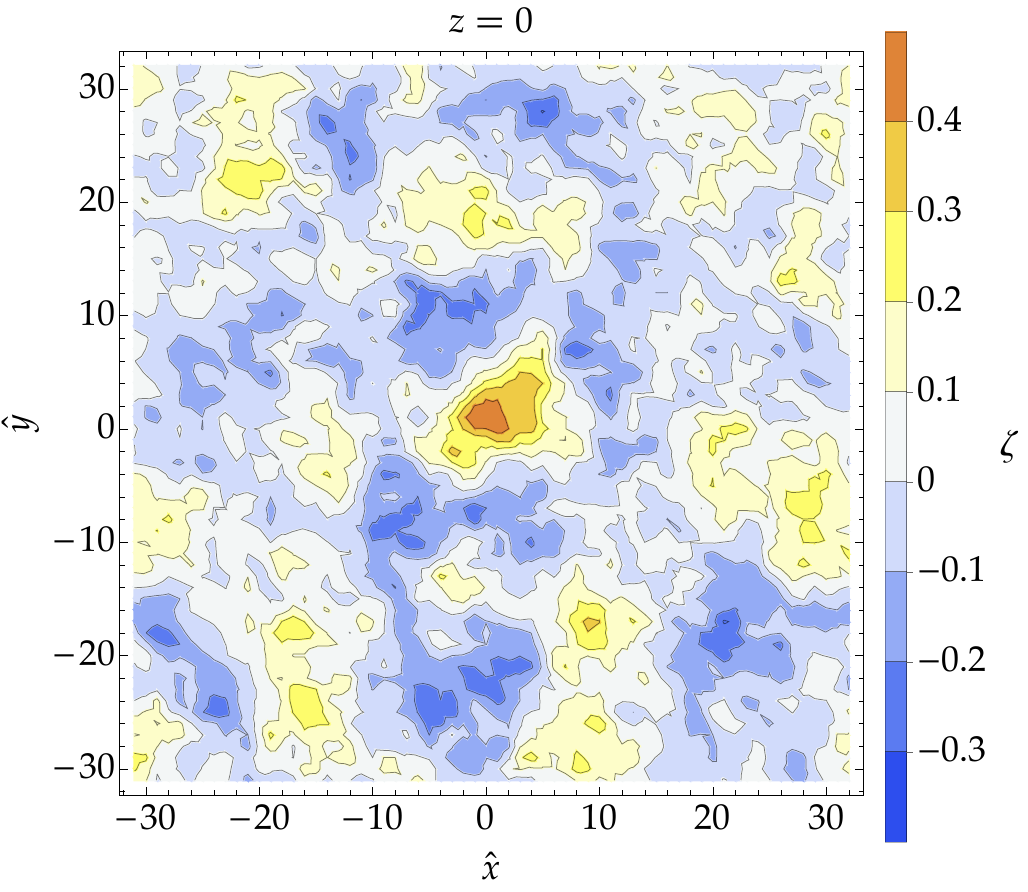}
        \end{minipage}
        \\\\
        \begin{minipage}{0.48\hsize}
            \centering
            \includegraphics[width=0.9\hsize]{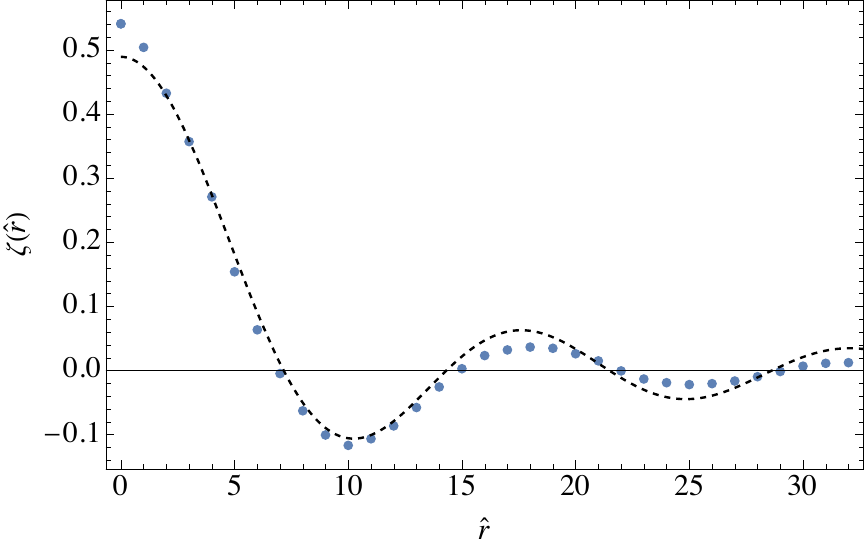}
        \end{minipage}&
        \begin{minipage}{0.48\hsize}
            \centering
            \includegraphics[width=0.9\hsize]{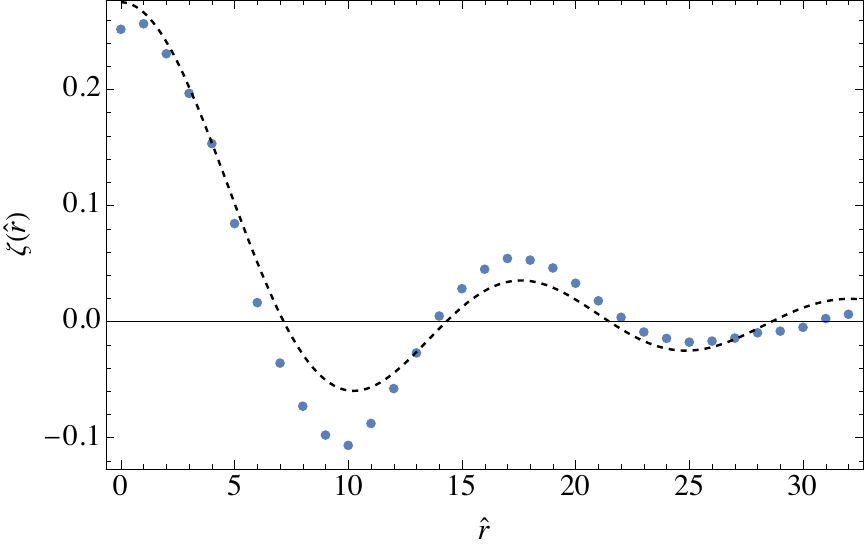}
        \end{minipage}\\\\
        \begin{minipage}{0.48\hsize}
            \centering
	        \includegraphics[width=0.95\hsize]{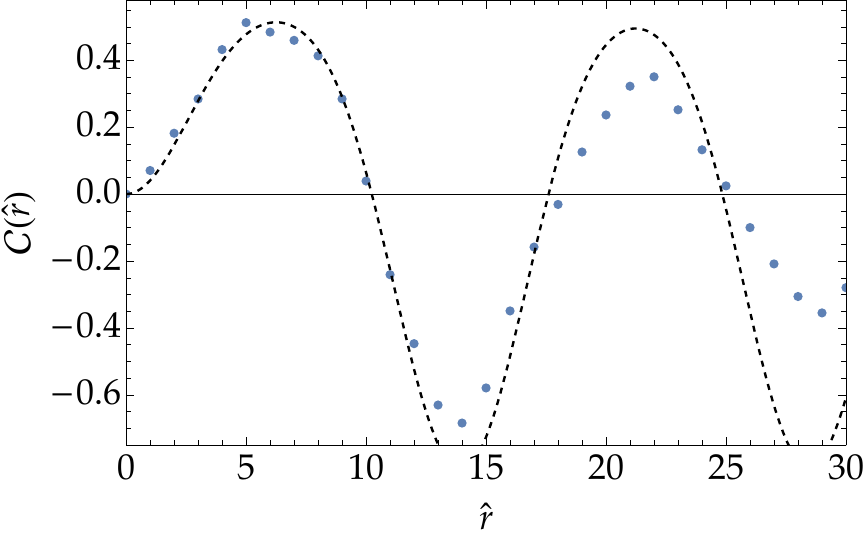}
        \end{minipage}&
        \begin{minipage}{0.48\hsize}
            \centering
            \includegraphics[width=0.95\hsize]{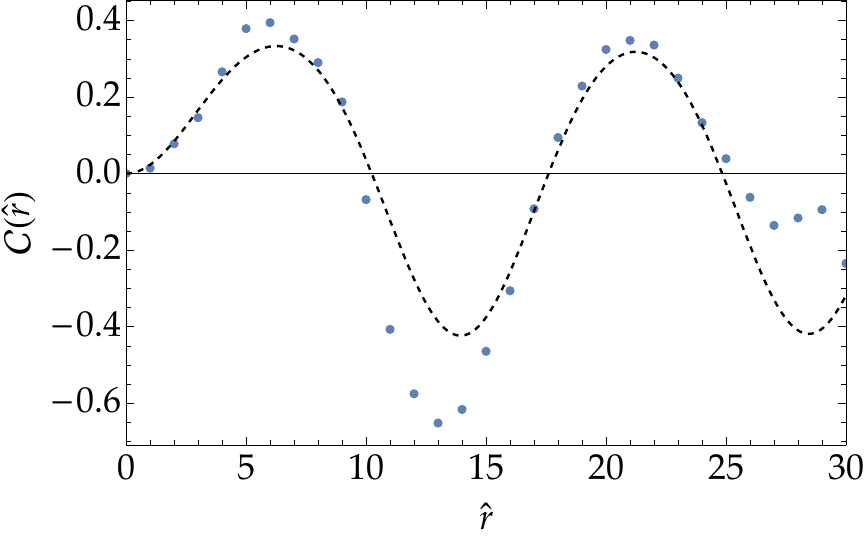}
        \end{minipage}\\
	\end{tabular}
	\caption{Samples of the contours of $\zeta$ for the $z=0$ slice (top), the corresponding radial profile $\zeta(\hat{r})$ (middle), and the compaction function $\calC(\hat{r})$ (bottom) for Starobinsky's linear model.
    The left panels correspond to a \ac{PBH}-forming realisation $\bar{\calC}_\um>2/5$, while the right panels are for a non-\ac{PBH}-forming one $\bar{\calC}_\um<2/5$. In the middle and bottom panels, the blue dots are the numerical results and the black-dashed lines represent the $\sinc$-profile fitting with $n_\sigma(N_\ub)$.
    }
	\label{fig: under0.4_over0.4}
\end{figure}

Let us see the results of the importance sampling in Starobinsky's linear model with the same parameters discussed in Sec.~\ref{sec: Starobinskys linear}.
We adopt the box size $L=10^{-12}\,\si{Mpc}$ and
$(b,N_\ub,\Delta N_\ub)=(20\sqrt{\Delta N_\ub},3.8,0.1)$ for the bias parameters in Eq.~\eqref{eq: bias}, where $N_\ub$ corresponds to the peak wave number $n_*=4.47$ of the power spectrum shown in Fig.~\ref{fig: StatStarobinsky} via the cutoff scale $n_\sigma(N_\ub)=\sigma\ee^{N_\ub}$.
In Fig.~\ref{fig: under0.4_over0.4}, we show two sample $xy$-slices at $z=0$ of the biased $\zeta$ contour map for a \ac{PBH}-forming $\bar{\calC}_\um>2/5$ realisation and not $\bar{\calC}_\um<2/5$. The corresponding spherical-average profile $\zeta(\tilde{r})$ and numerically obtained compaction function $\calC(\tilde{r})$~\eqref{eq: compaction} as functions of the normalised radius $\tilde{r}=\frac{N_L}{L}\abs{\bfx}$ are compared with the correspondings of the $\sinc$ profile~\eqref{eq: typical profile}. They are consistent around the peak for the \ac{PBH}-forming realisation as the power spectrum is well peaked around $n_\sigma(N_\ub)$ and the curvature perturbation is almost Gaussian.

\begin{figure}
	\centering
	\begin{tabular}{cc}
		\begin{minipage}{0.5\hsize}
			\centering
			\includegraphics[width=0.95\hsize]{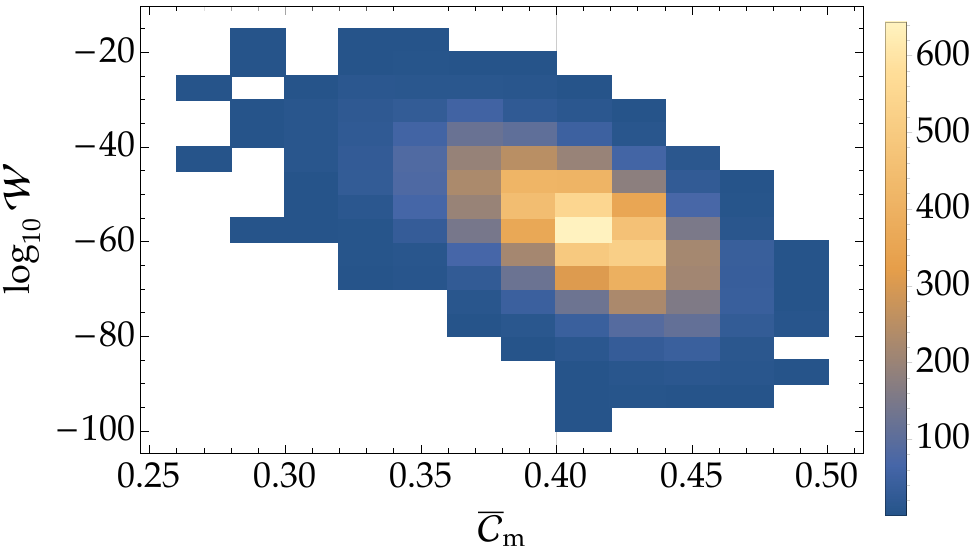}
		\end{minipage}&
		\begin{minipage}{0.5\hsize}
			\centering
			\includegraphics[width=0.95\hsize]{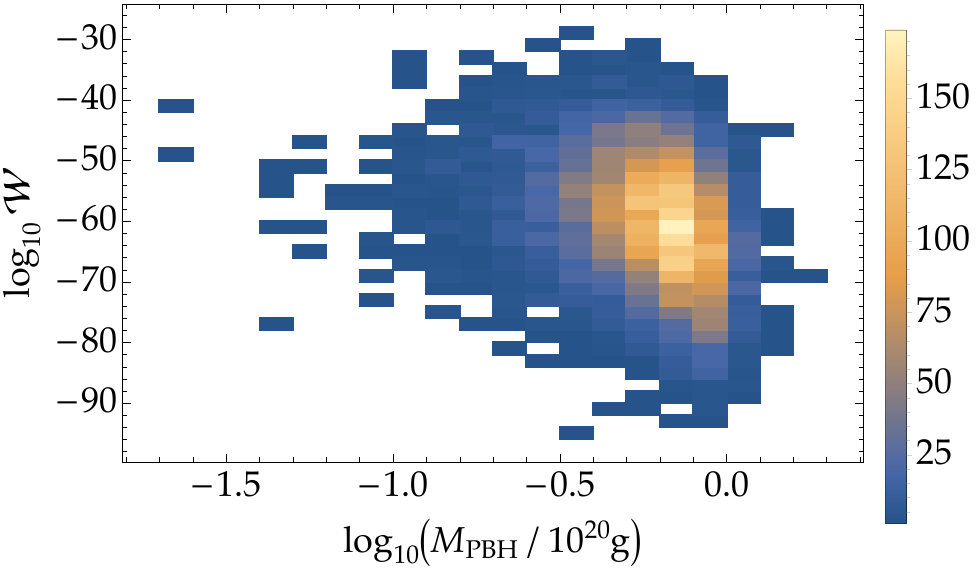}
		\end{minipage}
	\end{tabular}
	\caption{\emph{Left}: the density histogram of the average compaction function $\bar{\calC}_\um$~\eqref{eq: averagecompaction} and the weight function $\calW$~\eqref{eq: lnwN} of the 10000 samples for Starobinsky's linear model. Thanks to the bias, the \ac{PBH}-forming maps $\bar{\calC}_\um\simeq\bar{\calC}_\uth=2/5$ (shown by the vertical thin line) are intensively realised. The negative correlation between $\bar{\calC}_\um$ and $\calW$ indicates that larger overdensities are rarer as expected. 
	\emph{Right}: the same plot with respect to the \ac{PBH} mass $M_\PBH$~\eqref{eq: M_Horizon} and the weight for 6396 samples for which $\bar{\calC}_\um$ exceeds the threshold $\bar{\calC}_\uth=2/5$. }
	\label{fig: density histogram}
\end{figure}

\begin{figure}
	\centering
	\begin{tabular}{cc}
		\begin{minipage}{0.48\hsize}
			\centering
			\includegraphics[width=0.95\hsize]{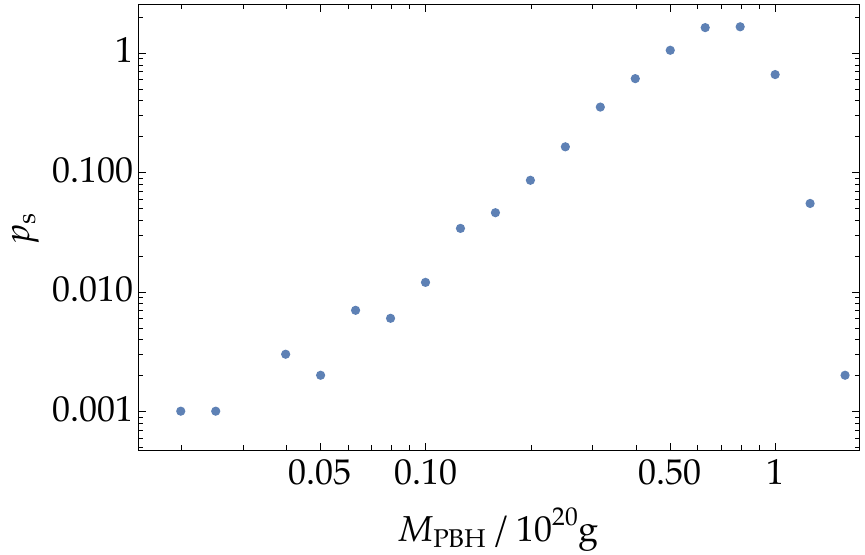}
		\end{minipage}&
		\begin{minipage}{0.48\hsize}
			\centering
			\includegraphics[width=0.95\hsize]{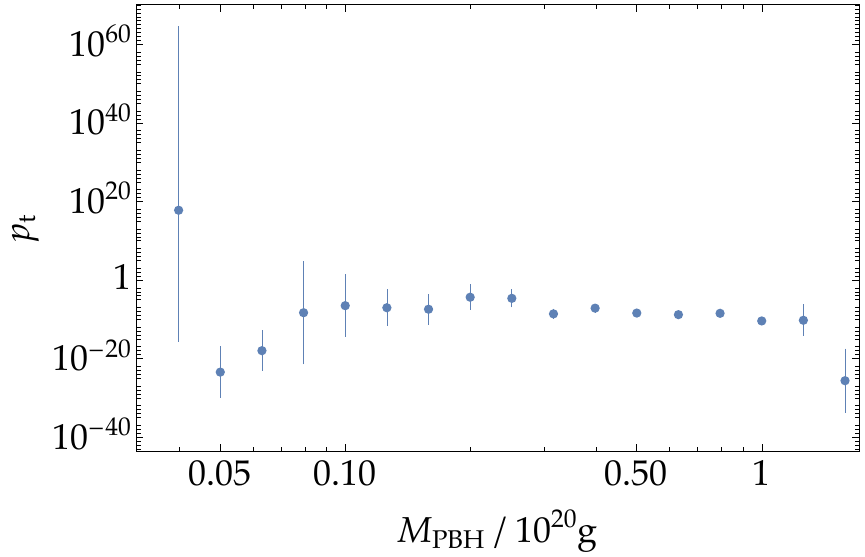}
		\end{minipage}
	\end{tabular}
	\caption{The sampling probability $p_\us$ (left) and the target probability $p_\ut$ (right) with respect to the \ac{PBH} mass.
    }
	\label{fig: probabilities}
\end{figure}

\begin{figure}
	\centering
	\includegraphics[width=0.7\hsize]{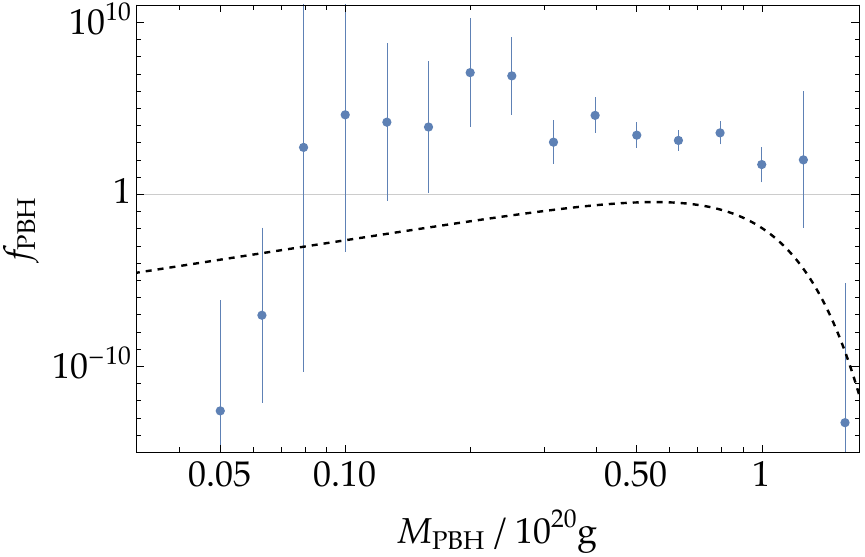}
	\caption{The total \ac{PBH} abundance $f_\PBH$ for Starobinsky's linear model.
    We adopt $g_*=106.75$ and $\Omega_\DM h^2=0.12$~\cite{Planck:2018vyg}.
    The blue dots with error bars are the numerical results, while the black-dashed line is the estimation of the peak theory in the monochromatic assumption~\eqref{eq: monochromatic power} at $n_\sigma(N_\ub)\simeq4.47$ with the amplitude $A_\us=\num{5.054e-3}$.
    }
	\label{fig: f_PBH_Sta}
\end{figure}

The left panel of Fig.~\ref{fig: density histogram} shows the density histogram in the average compaction function $\bar{\calC}_\um$~\eqref{eq: averagecompaction} and the weight function $\calW$~\eqref{eq: lnwN}.
Their negative correlation indicates that larger overdensities are rarer as expected and the choice of the bias function~\eqref{eq: bias} works well.
For \ac{PBH}-forming maps $\bar{\calC}_\um>2/5$, the corresponding \ac{PBH} mass $M_\PBH$ can be calculated by Eq.~\eqref{eq: M_Horizon} with the computed $\bar{\calC}_\um$ and $R_\um$ and then one finds the density histogram in $M_\PBH$ and $\calW$ as shown in the right panel of Fig.~\ref{fig: density histogram}.
Through the lognormal estimator~\eqref{eq: lognormal estimator}, it converts the sampling probability $p_\us$ by the histogram~\eqref{eq: ps} (the left panel of Fig.~\ref{fig: probabilities}) into the target probability $p_\ut$ shown in the right panel of Fig.~\ref{fig: probabilities}.
It is directly followed by the current \ac{PBH} mass function $f_\PBH$ via the formula~\eqref{eq: fPBH in importance sampling}. Our final result is shown in Fig.~\ref{fig: f_PBH_Sta} with the prediction of the peak theory as a comparison.
Here, the peak theory calculation is done by first fitting the peak of the power spectrum by the log-normal function (the orange-dotted line in the left panel of Fig.~\ref{fig: StatStarobinsky}) and approximating it by the Dirac delta function~\eqref{eq: monochromatic power} with the same amplitude $A_\us=\num{5.054e-3}$.
Despite their completely independent approaches, the amplitude and scale dependence of their independent result are roughly consistent with the known analytical calculation, indicating that \ac{STOLAS} with the importance sampling technique works.
The quantitative difference would be caused by the monochromatic power spectrum assumption~\eqref{eq: monochromatic power} in the peak theory or the low spatial resolution of \ac{STOLAS}, which we leave for future work.

\begin{figure}
    \centering
    \begin{tabular}{c}
        \begin{minipage}{0.55\hsize}
            \centering
            \includegraphics[width=0.95\hsize]{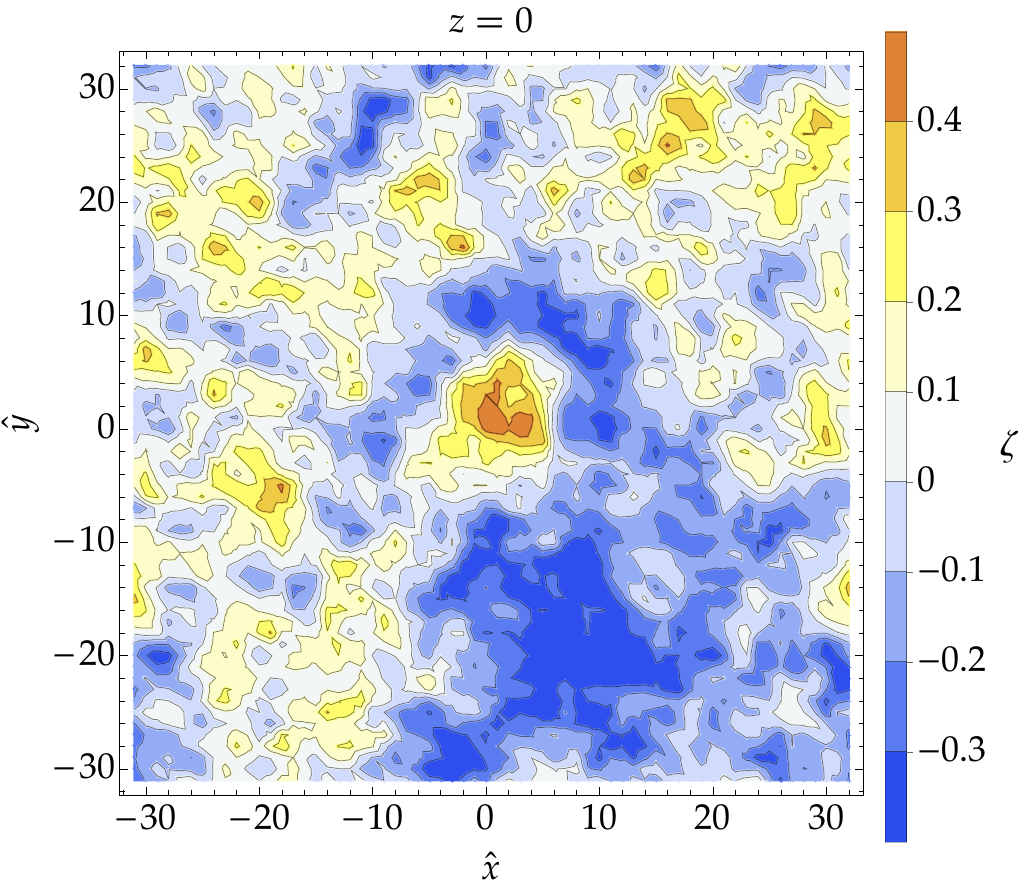}
        \end{minipage}
        \begin{minipage}{0.35\hsize}
            \centering
            \includegraphics[width=0.9\hsize]{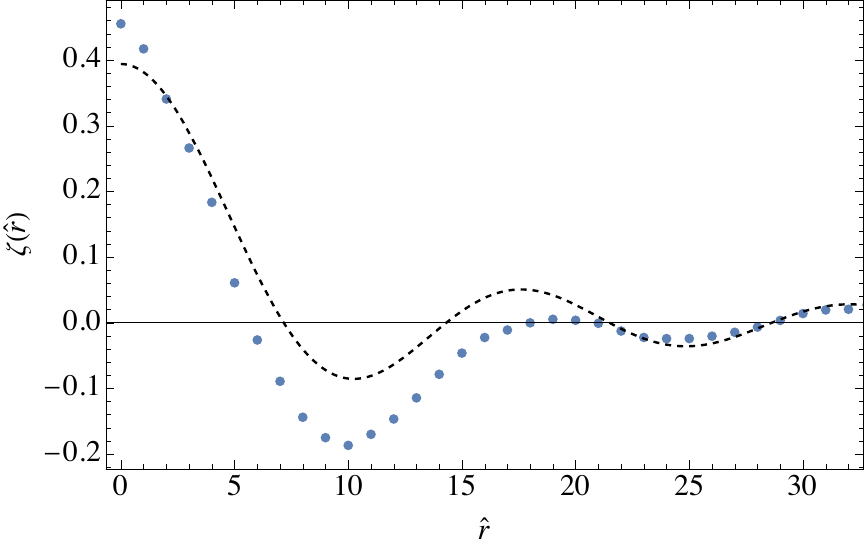}\\
            \includegraphics[width=0.95\hsize]{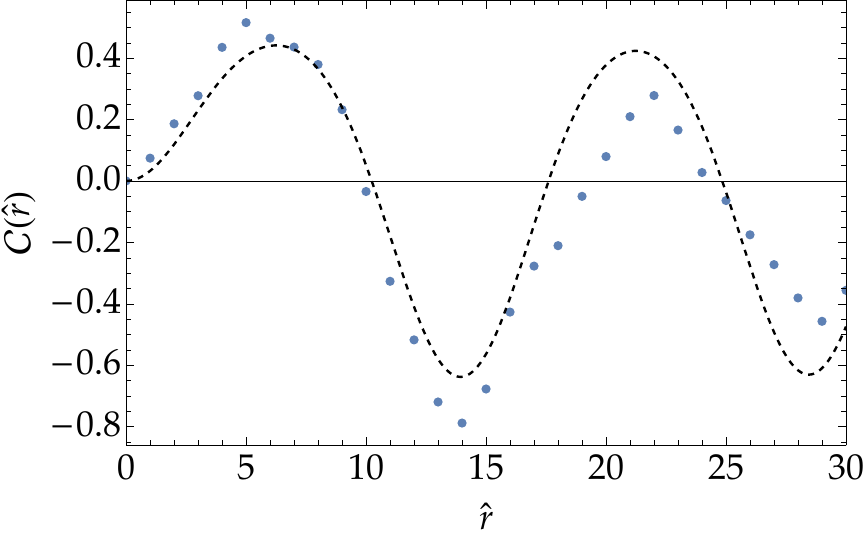}
        \end{minipage}\\\\
        \begin{minipage}{0.45\hsize}
            \centering
            \includegraphics[width=0.95\hsize]{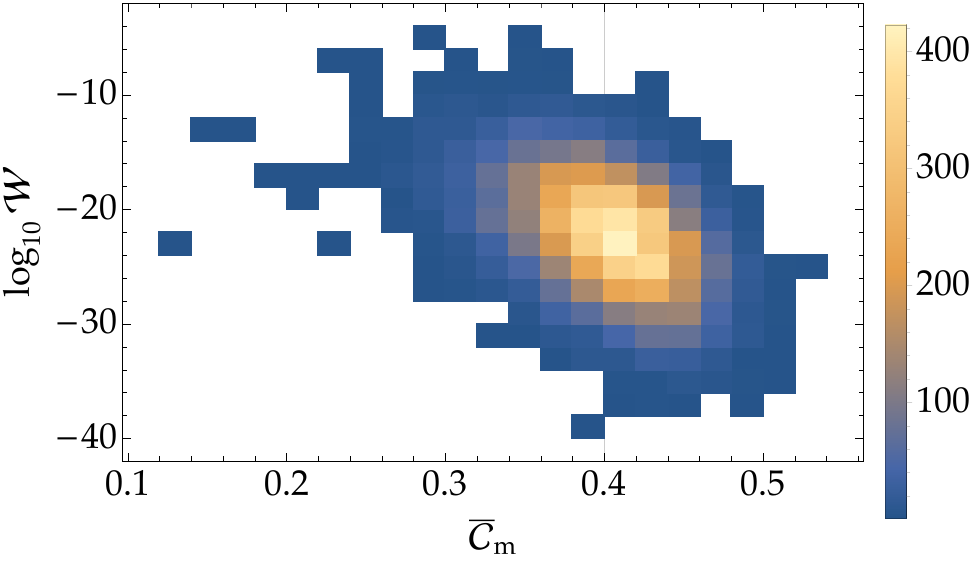}
        \end{minipage}
        \begin{minipage}{0.45\hsize}
		  \centering
			\includegraphics[width=0.95\hsize]{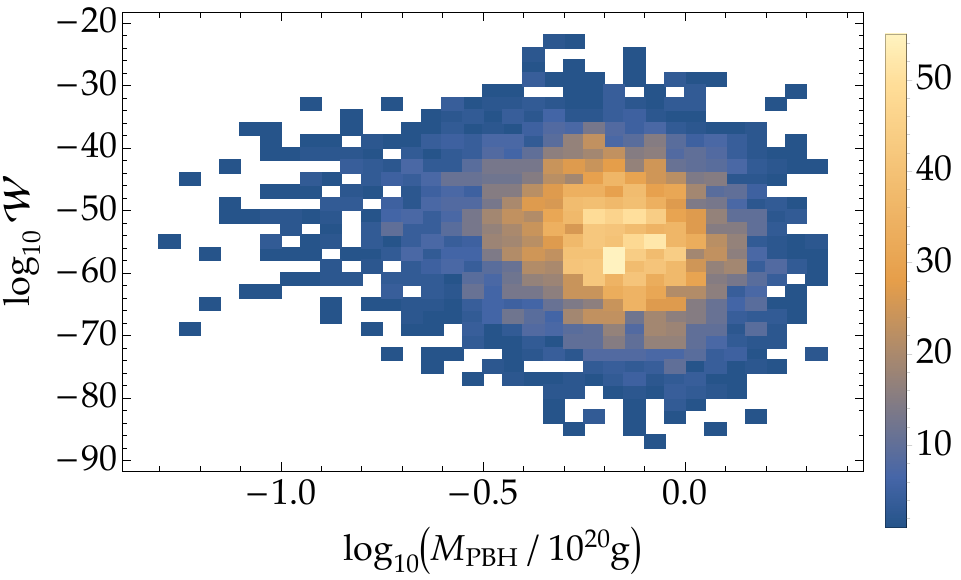}
		\end{minipage}\\\\
        \begin{minipage}{\hsize}
            \centering
            \includegraphics[width=0.7\hsize]{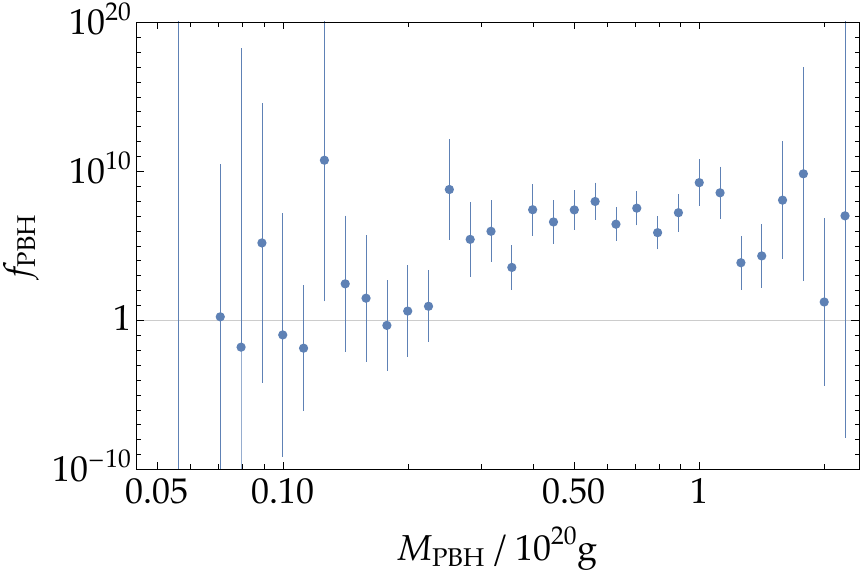}
        \end{minipage}
    \end{tabular}
    \caption{The same plots as Figs.~\ref{fig: under0.4_over0.4}--\ref{fig: f_PBH_Sta} for chaotic inflation. The same noise map is used for the top panels as the left panels of Fig.~\ref{fig: under0.4_over0.4}.}
    \label{fig: statistics_chaotic}
\end{figure}

\begin{figure}
	\centering
	\begin{tabular}{cc}
        \begin{minipage}{0.48\hsize}
			\centering
			\includegraphics[width=0.95\hsize]{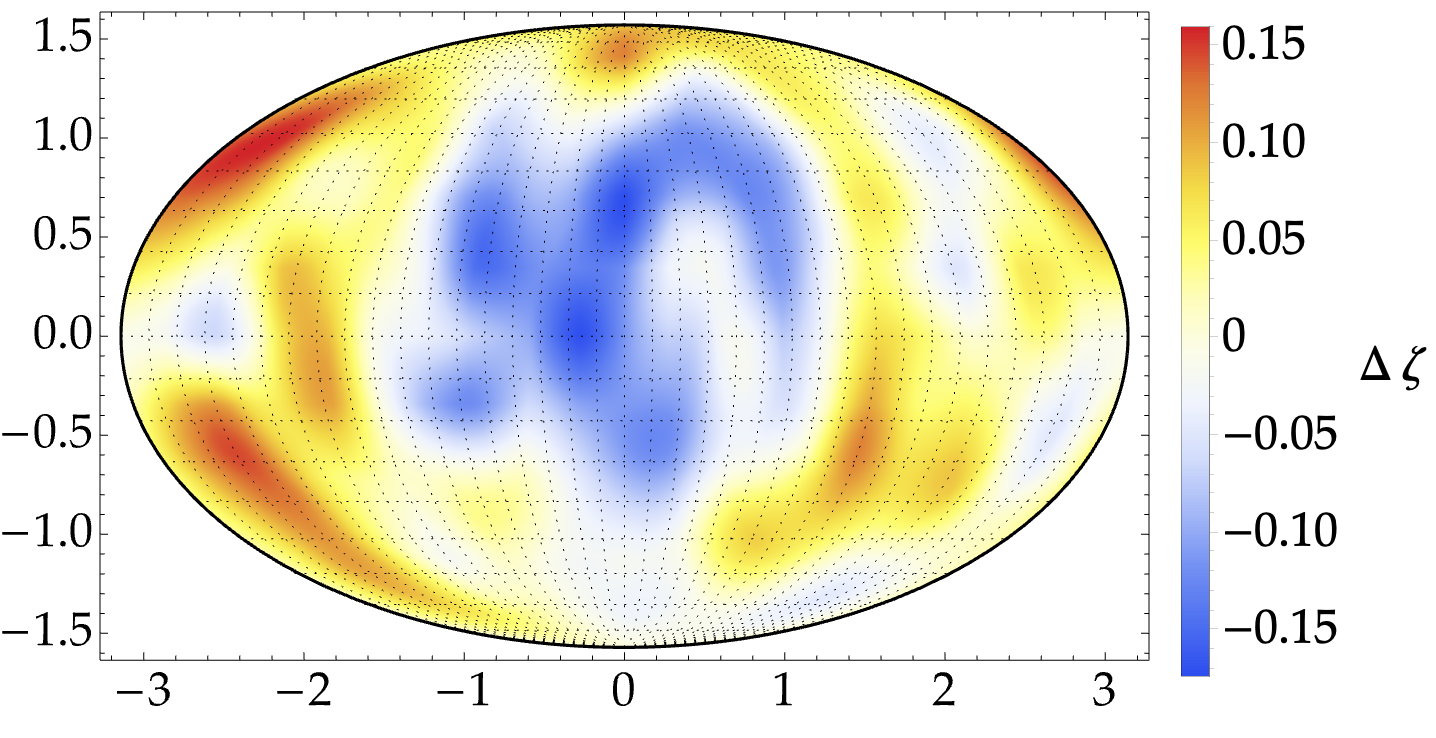}
		\end{minipage}&
		\begin{minipage}{0.48\hsize}
			\centering
			\includegraphics[width=0.95\hsize]{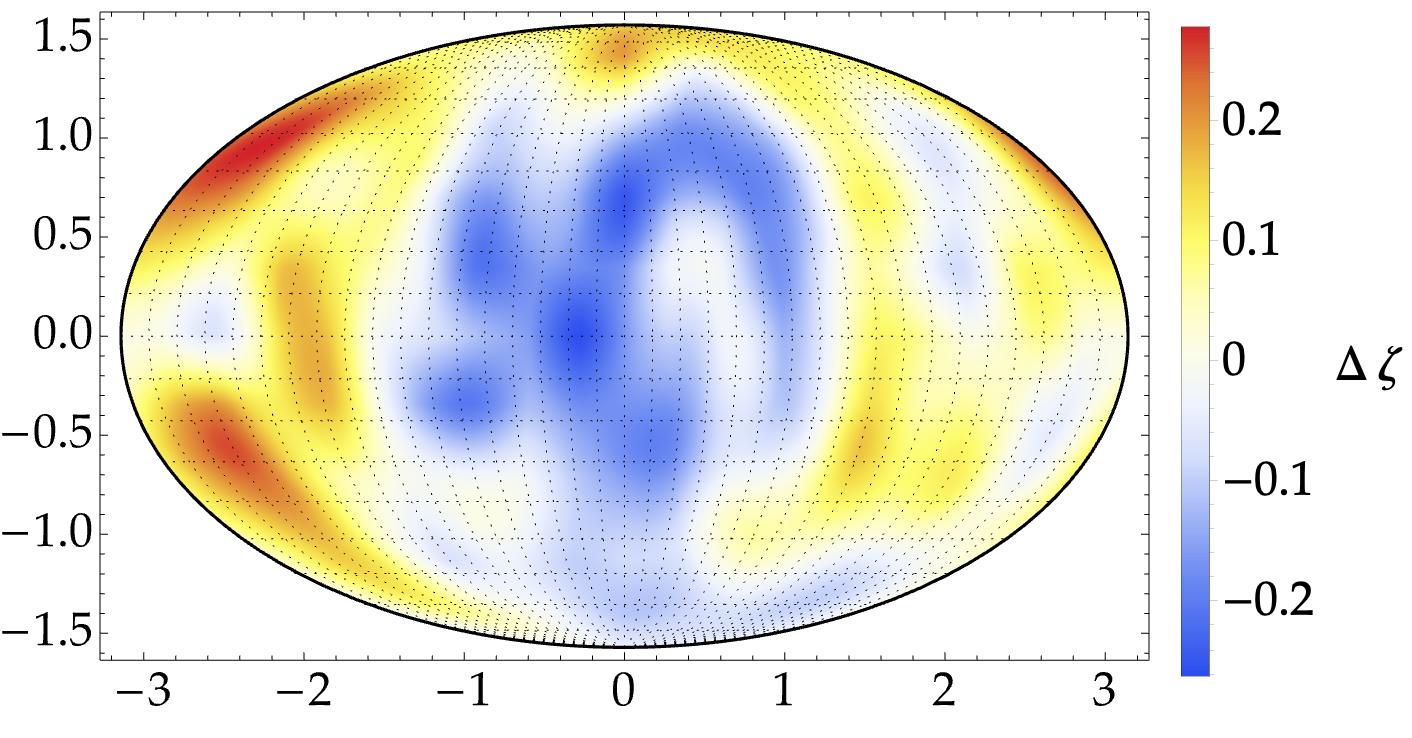}
		\end{minipage}
	\end{tabular}
	\caption{Examples of the angular dependence of the curvature perturbation around the spherical average, $\Delta\zeta$, at $\hat{r}=3.6$ corresponding to the quarter wave $N_L/(4n_\sigma(N_\ub))$. The left panel is for Starobinsky's linear model while the right is for chaotic inflation. 
    We use the same noise map.
    One finds that the broader power spectrum in chaotic inflation causes a more non-spherical overdensity, which may affect the \ac{PBH} formation criterion.}
	\label{fig: Moellweide}
\end{figure}

The results for chaotic inflation are shown in Fig.~\ref{fig: statistics_chaotic}.\footnote{The peak theory result is not shown because the calculation for the nearly scale-invariant or broad power spectrum with the average compaction criterion has not been given in the literature.}
One can find a broad feature of the mass function reflecting the nearly scale-invariant power spectrum shown in Fig.~\ref{fig:StatChaotic} though the bias is introduced only for a specific scale $\sim n_\sigma(N_\ub)$. It is enabled by the randomly added noise $\Delta W(N,\bfx)$.
One can extract $f_\PBH$ for different masses more accurately by changing the bias time $N_\ub$.

We mention a caveat in this result. As shown in Fig.~\ref{fig: Moellweide}, chaotic inflation generates more non-spherical overdensities due to its broad power spectrum. In fact, we calculated
the angular power spectrum
\bae{
    C_\ell=\frac{1}{2\ell+1}\sum_{m=-\ell}^{+\ell}\abs{a_{\ell m}}^2 \qc
    a_{\ell m}=\int \zeta(\Omega)Y_\ell^m(\Omega)\dd{\Omega},
}
where $Y_\ell^m(\Omega)$ is the spherical harmonics and found 
$C_1=(1.095\pm0.009)\times10^{-2}$ and 
$C_2=(3.123\pm0.020)\times10^{-3}$ for Starobinsky's linear model while $C_1=(2.112 \pm 0.017)\times 10^{-2}$ and $C_2=(6.415\pm0.040)\times 10^{-3}$
for chaotic inflation at the quarter wave  $\hat{r}=N_L/(4n_\sigma(N_\ub))\simeq3.6$ where the standard errors are estimated with $10,000$ samples.
Large non-sphericity may cause a wrong judgement of \ac{PBH} formation with the spherical compaction function criterion as indicated by the worse fitting of the $\sinc$ profile or the less correlation between $\calW$ and $\bar{\calC}_\um$ in Fig.~\ref{fig: statistics_chaotic}.
We leave it for future work.

\section{Conclusions}\label{sec: Conclusions}

In this paper, we proposed the C++ package, \textsc{\acf{STOLAS}}, of cosmic inflation. It simulates the dynamics of the inflaton fields in the manner of the stochastic formalism, the effective theory of superHubble fields, and calculates the observable curvature perturbation according to the $\delta N$ formalism.
In Sec.~\ref{sec: Stochastic lattice simulation}, we described the implementation of \ac{STOLAS} by discretising the stochastic formalism.
In Sec.~\ref{sec: Primary statistics}, we calculated primary statistics such as the power spectrum and the non-linearity parameter of the curvature perturbation in two toy models: chaotic inflation and Starobinsky's linear-potential inflation. 
The results (Figs.~\ref{fig:StatChaotic} and \ref{fig: StatStarobinsky}) show the consistency of \ac{STOLAS} with the standard perturbation theory in the perturbative observables.
As a further application of \ac{STOLAS} beyond the perturbative quantities, we calculated the abundance of \acp{PBH} in the two models in Sec.~\ref{sec: Importance sampling} by introducing the technique of the importance sampling, which can efficiently sample rare events such as \acp{PBH}. 
For the first time, we had success in directly sampling the \ac{PBH} abundance $f_\PBH$ in a full numerical way as shown in Figs.~\ref{fig: f_PBH_Sta} and \ref{fig: statistics_chaotic}.
The quantitative inconsistency between \ac{STOLAS} and the analytic result in the peak theory may be caused by the monochromatic power spectrum approximation in the peak theory and/or the low spatial resolution of \ac{STOLAS}. 
We leave a detailed analysis of this inconsistency for future work.

The simulation reveals that the non-sphericity of the overdensity tends to be relatively large for a broader power spectrum represented by chaotic inflation (Fig.~\ref{fig: Moellweide}). It will hence be interesting to investigate the \ac{PBH} formation criteria for a non-spherical overdensity in numerical relativity (see, e.g., Ref.~\cite{Yoo:2020lmg} for such an attempt).
Also, while the non-Gaussianity of the curvature perturbation is small in both models we consider, particularly in Starobinsky's linear model because the end of \ac{USR} phase is smoothly connected to the second slow-roll phase in our setup, one may introduce a sharp transition from \ac{USR} to slow-roll which is known to cause the \emph{exponential-tail} feature in the curvature perturbation.
This non-perturbative feature can significantly enhance the \ac{PBH} abundance and therefore be a good target of the investigation with \ac{STOLAS}.
Multifield extensions such as hybrid inflation (see, e.g., Refs.~\cite{Kawasaki:2015ppx,Tada:2023fvd,Tada:2024ckk}) are also attractive as they predict non-Gaussian curvature perturbations in general (see, e.g., Refs.~\cite{Lyth:2010zq,Lyth:2012yp,Bugaev:2011qt,Bugaev:2011wy} in the context of hybrid inflation).

\ac{STOLAS}'s application is not limited to the \ac{PBH}.
The map of the curvature perturbations obtained in \ac{STOLAS} could be used as the initial condition of the structure formation simulation such as $N$-body simulations.
Since it does not require translating the information of the curvature perturbation into small numbers of statistics such as the power or bispectrum, one can extract the full non-perturbative features of the curvature perturbation embedded into the large-scale structure (see, e.g., Ref.~\cite{Ezquiaga:2022qpw} for the effect of the exponential tail to the number of massive clusters).

\acknowledgments

We are grateful to Takashi Hiramatsu, Naoya Kitajima, and Shuichiro Yokoyama for their helpful discussions, and to Angelo Caravano, Diego Cruces, Albert Escriv\`a, Cristiano Germani, Rhodai Kawaguchi, Eiichiro Komatsu, Lucas Pinol, Koki Tokeshi, and Vincent Vennin, Edoardo Vitagliano for their useful comments on the draft. 
We also thank the organisers and participants of \emph{Observational Cosmology Summer Workshop in 2022}, where the key idea of the work was given.
This work is supported by JSPS KAKENHI Grants
No.~JP23KJ2007 (T.M.),
No.~JP21K13918 (Y.T.),
and No.~JP24K07047 (Y.T.).

\appendix

\section{Derivation of the power spectrum of the scalar field at \acs{NLO} in slow-roll}\label{sec: calPphi}

In this appendix, we derive the power spectrum of the scalar field up to the \ac{NLO} in the slow-roll expansion.
Note that we ignore the noise term throughout this appendix.
Following Ref.~\cite{Auclair:2022yxs}, the slow-roll parameters are given by
\begin{align}
  \epsilon_{1} = - \frac{\dot{H}}{H^2},
  \quad
  \epsilon_{2} = \frac{\dot{\epsilon}_{1}}{H \epsilon_{1}}.
\end{align}
The relationship between the conformal time and the comoving horizon is calculated as
\begin{align}
  a H \simeq -\frac{1}{\tau}(1 + \epsilon_{1}).
\end{align}
The Mukhanov--Sasaki equation for the canonicalised inflaton perturbation $v=a\delta\phi$ is given by
\begin{align}
  \pdv[2]{v}{\tau}
  + \qty(k^2 - \frac{1}{z} \pdv[2]{z}{\tau}) v
  =0,
  \label{eq: MSeq}
\end{align}
where $z=a\dot{\phi}/H$.
We can express the last term by using slow-roll parameters as
\begin{align}
  \frac{1}{z} \pdv[2]{z}{\tau}
  &\simeq \frac{1}{\tau^2}\qty(2 + 3 \epsilon_{1} + \frac{3}{2}\epsilon_{2}).
\end{align}
Supposing the slow-roll parameters are almost constant up to \ac{NLO}, Eq.~\eqref{eq: MSeq} can be solved analytically as 
\begin{align}
  v (\tau) &= \sqrt{\frac{\pi}{4k}} \ee^{i\pi (\nu+1/2)/2}
  \sqrt{-k\tau} H_{\nu}^{(1)}(-k\tau),
\end{align}
where $H_{\nu}^{(1)}(x)$ is the Hankel Function of the First Kind, and the parameter $\nu$ is given by
\begin{align}
  \nu \simeq \frac{3}{2} + \epsilon_{1} + \frac{1}{2}\epsilon_{2}.
\end{align}
Here, we chose the integration constant consistent with the adiabatic vacuum in the subHubble limit.
The asymptotic form of the Hankel function in the superHubble limit is given by
\begin{align}
  H_{\nu}^{(1)}(x) \to 
  -i \frac{\Gamma (\nu)}{\pi} \qty(\frac{2}{x})^2,
\end{align}
where the Gamma function $\Gamma(\nu)$ is expanded as
\begin{align}
  \Gamma (\nu)
  &\simeq \Gamma \qty(\frac{3}{2})
  \qty[1 + \qty(\epsilon_{1} + \frac{1}{2}\epsilon_{2}) \qty(2-\gamma-2\ln{2})]
\end{align}
with the Euler's constant $\gamma$.
Using these relations, the power spectrum of $\phi$ is computed as
\begin{align}
  {\cal P}_{\phi} (\sigma a \sfH) &= \frac{k^3}{2 \pi a^2}
  \frac{-\tau}{4}
  \abs{H_{\nu}^{(1)}(-k\tau)}^2
  \notag\\
  &= \qty(\frac{H}{2\pi})^2
  \qty[1 + 2\epsilon_{1} \qty(1-\gamma-2\ln{2})
  + \epsilon_{2} \qty(2-\gamma-2\ln{2})]
  \qty(\frac{\sigma \sfH}{2H})^{-2\epsilon_{1}-\epsilon_{2}}.
\end{align}
One can change the slow-roll parameters to the potential form with the use of the relation
\begin{align}
  \epv = \frac{\Mpl^2}{2} \frac{V'}{V} \simeq \epsilon_{1},
  \quad
  \hv = \Mpl^2 \frac{V''}{V} \simeq 2\epsilon_{1} - \frac{1}{2}\epsilon_{2}.
\end{align}
Then, the power spectrum is written as
\begin{align}
  {\cal P}_{\phi} (\sigma a \sfH) &= 
  \qty(\frac{H}{2\pi})^2
  \qty[1 + \epv \qty(10 - 6 \gamma -12\ln{2}) - 2 \hv\qty(2-\gamma-2\ln{2}) ]
  \qty(\frac{\sigma \sfH}{2 H})^{- 6 \epv + 2\hv}.
\end{align}
Practically, this form is more useful because the second time derivative in $\epsilon_2$ is not well defined in the stochastic system.

\bibliographystyle{JHEP_modified}
\bibliography{main}
\end{document}